\documentclass[12pt]{article}
\usepackage[onehalfspacing]{setspace}
\usepackage[english]{babel}
\usepackage[T1]{fontenc}
\usepackage{times}
\usepackage{color}
\usepackage[left = 25mm, right = 25mm, bottom = 20mm, top = 25mm]{geometry}
\usepackage{amsmath}
\usepackage{amsfonts}
\usepackage{amssymb}
\usepackage{amsthm}
\usepackage{amscd}
\usepackage{natbib}
\usepackage{latexsym}
\usepackage{multirow}
\usepackage{bm}
\usepackage{bbm}
\usepackage{enumerate}
\usepackage{verbatim}
\usepackage{graphicx}
\usepackage{titlesec}
\usepackage{hyperref}
\usepackage{float}
\usepackage{rotating}
\usepackage{moreverb,url}
\usepackage{bibentry}
\usepackage{psfrag,epsf,color}
\usepackage{authblk}
\usepackage{microtype} 

\allowdisplaybreaks[4]

\setcitestyle{authoryear, round, comma}

\newtheorem{thm}{Theorem}

\newcommand{\mf}[1]{{\bf #1}}
\newcommand{\mfg}[1]{\boldsymbol{#1}}

\newcommand{\vfg}[1]{\boldsymbol{#1}}

\newcommand{\MatrFormat}[1]{{\bf #1}}
\newcommand{\MatrFormatGreek}[1]{\boldsymbol{#1}}
\newcommand{\VecFormat}[1]{{\bf #1}}
\newcommand{\VecFormatGreek}[1]{\boldsymbol{#1}}

\newcommand{\refmath}[1]{(\ref{#1})}
\newcommand{\refmathwhite}[1]{(W\ref{#1})}	
\newcommand{\refmathanc}[1]{(GA\ref{#1})}	
\newcommand{\AvgCovLine}{N^{-1}(\mf{X}'\mf{X})}
\newcommand{\AvgCovFrac}{\frac{\mf{X}'\mf{X}}{N}}
\newcommand{\best}{\vfg{\hat{\beta}}}
\newcommand{\bestwb}{\best^{\ast}}

\newcommand{\SigmaFull}{\mfg{\Sigma}_N}			
\newcommand{\SigmaWhiteFull}{\mfg{\hat{\Sigma}}_N}	

\newcommand{\SigmaPart}{\tilde{\boldsymbol{\Sigma}}}	
\newcommand{\SigmaWild}{\hat{\boldsymbol{\Sigma}}^{\ast}}						

\newcommand{\XMatr}{{\bf X}}
\newcommand{\condbar}{\,\big|\,}	

\defcitealias{IchE9}{ICH, 1998}
\defcitealias{Ema07}{EMA, 2007}
\defcitealias{Ema15}{EMA, 2015}
\defcitealias{Fda10}{FDA, 2010}
\defcitealias{Fda16}{FDA, 2016}
\defcitealias{Fda18}{FDA, 2018}

\begin{document}

\noindent
{\centering
\Large{{\bf Small-sample performance and underlying assumptions of a bootstrap-based inference method for a general analysis of covariance model with possibly heteroskedastic and nonnormal errors}}
}
\vspace*{1.5cm}

\noindent
{\large
\textbf{Authors} \\
\noindent
Georg Zimmermann (Department of Mathematics, Paris-Lodron-University of Salzburg, Salzburg, Austria)\\
\noindent 
Markus Pauly (Department of Mathematics, Ulm University, Ulm, Germany)\\
\noindent
Arne C. Bathke (Department of Mathematics, Paris-Lodron-University of Salzburg, Salzburg, Austria)
}

\vspace*{1.0cm}

\noindent
\textbf{Abstract}\\
\noindent
It is well known that the standard F test is severely affected by heteroskedasticity in unbalanced analysis of covariance (ANCOVA) models. Currently available potential remedies for such a scenario are based on heteroskedasticity-consistent covariance matrix estimation (HCCME). However, the HCCME approach tends to be liberal in small samples. Therefore, in the present manuscript, we propose a combination of HCCME and a wild bootstrap technique, with the aim of improving the small-sample performance. We precisely state a set of assumptions for the general ANCOVA model and discuss their practical interpretation in detail, since this issue may have been somewhat neglected in applied research so far. We prove that these assumptions are sufficient to ensure the asymptotic validity of the combined HCCME-wild bootstrap ANCOVA. The results of our simulation study indicate that our proposed test remedies the problems of the ANCOVA F test and its heteroskedasticity-consistent alternatives in small to moderate sample size scenarios. Our test only requires very mild conditions, thus being applicable in a broad range of real-life settings, as illustrated by the detailed discussion of a dataset from preclinical research on spinal cord injury. Our proposed method is ready-to-use and allows for valid hypothesis testing in frequently encountered settings (\textit{e.g.}, comparing group means while adjusting for baseline measurements in a randomized controlled clinical trial).

\vspace*{0.5cm}
\noindent
\textbf{Keywords}\\
\noindent
Heteroskedasticity-consistent covariance matrix
estimator, Analysis of covariance, Rare disease, Resampling, Small sample, Wild bootstrap


\newpage
\section{Introduction}\label{Intro}

\subsection{The analysis of covariance model and its assumptions}
Consider the frequently encountered situation where several groups of subjects are being compared
with respect to a continuous outcome variable. For the statistical comparison of the group
means, the analysis of variance (ANOVA) is often used. However, in many instances, it is reasonable to account
for one or several covariates, such as baseline measurements or variables which are thought to be related to the outcome. A recently published EMA guideline for clinical trials recommends adjusting for any variable which is at least moderately associated with the primary outcome \citepalias{Ema15}. 
For this purpose, the analysis of covariance (ANCOVA) is an appropriate tool, which is used
with the aim of increasing the inferential power, and reducing bias and variance of the effect estimators \citep{Hui}. The ANCOVA has been applied in many research disciplines, ranging from 
studies about Alzheimer's disease \citep{Bos} to pharmaceutical issues \citep{Lu14}, educational \citep{keselman1998statistical} and fishery research \citep{Mis}, just to name a few. 

There have been controversial discussions concerning the appropriate
use and interpretation of ANCOVA in various settings \citep{Ada, keselman1998statistical, Ber, Owe, Poc, Sen}. Apart from that, it is well known that, as an inference method, it also relies heavily on the assumptions of homoskedasticity and normality
of the errors. It has been shown in simulation studies that the violation of one or more of these
assumptions can seriously affect the ANCOVA F test in terms of maintenance of the prespecified
type I error level and power \citep{glass1972consequences,Hui}.

Many of the proposed solutions to tackle this problem can be grouped into one of the following two approaches. Namely, on the one hand, some authors essentially kept the fully parametric ANCOVA model, but tried to derive test
statistics which are more robust against violations of the above-mentioned assumptions. This approach had already
been considered several decades ago \citep{Ash} and has recently been enriched by different bootstrap variations \citep{Sad}. 
On the other hand, leaving the parametric model in  favour of a nonparametric approach has received attention. In this regard, both introductory papers explaining
the proper application of ANCOVA to real-life data \citep{Koc, Tan, lesaffre2003note}, as well as more theoretic contributions \citep{Bat, Tsi, Tha, Cha, Fan17} have been published. 
However, the latter approaches still impose some restrictions regarding either the number of groups or the number of covariates. 
Moreover, only moderate to large sample size scenarios have been considered so far, 
with a minimum number of 40 subjects per group \citep{Cha}. 
Thus, the small-sample performance of those methods remains unknown. However, group sizes below that level are  encountered quite frequently and there is a need for precise statistical methods for such situations. Examples cover data from preclinical research as well as studies on rather rare diseases ({\it e.g.}, spinal cord injury or multiple sclerosis). In one single paper, the authors indeed examined the small-sample performance of the tests they proposed with respect to the 
maintenance of the nominal $\alpha$ level. 
They found that finite-sample properties also depended on the number of groups \citep{Bat}. 
Likewise, the enriched parametric ANCOVA approach lacks sufficient evidence regarding its 
performance in finite-sample situations. 
For example, in simulation studies, only the homoskedasticity assumption was relaxed, but the normality assumption was maintained \citep{Sad}.

\subsection{Heteroskedasticity-consistent covariance matrix estimation in regression analysis}
Relaxing the model assumptions has been an important focus of research in the field of
regression analysis for some decades, too. 
Especially with regard to heteroskedasticity of the errors, an important contribution 
was White's heteroskedasticity-consistent covariance matrix estimator (HCCME),
and the derivation of the asymptotic distributions of its corresponding
test statistics \citep{Whi}. In medical studies, this approach has been used to some extent, too, both 
in applied branches such as diffusion tensor imaging \citep{Zhu} and genome-wide microarray
analyses \citep{Bar}, as well as in more methodological papers \citep{Kim, Jud}. 
However, the statement that HCCMEs are hardly 
known outside the statistical audience \citep{Hay} still appears to be a fair assessment. 
Moreover, the methods discussed so far either lack sufficient generality, 
maintaining strong assumptions like the normality of the errors \citep{Kim}, 
or they do not consider the assumptions underlying HCCMEs at all \citep{Hay, Zhu, Bar, Jud}. In addition to that, results of simulation studies conducted in the context of linear regression indicate that the classical asymptotic HCCME-based tests tend to be liberal \citep{Mac10}. However, it remains an open question whether this holds true for the ANCOVA model, too. Moreover, proofs are hardly provided, except in one publication, where the authors nevertheless kept the restrictive assumption of a symmetric error distribution, yet providing some empirical evidence that their approach might work in a more general setting, too \citep{Dav}.
\indent
\subsection{Objectives}
The aim of our work is twofold: Firstly, we state a set of assumptions for the general
ANCOVA model and prove that they are sufficient for the White HCCME approach \citep{Whi} to 
be applied. Henceforth, we will refer to this approach as well as to the associated 
test statistic by the term White-ANCOVA. Moreover, we discuss what these assumptions actually mean in practice and 
demonstrate that our approach covers a broad variety of designs which are frequently used
in applied research, including multi-way layouts and models with hierarchically nested factors. 
Hopefully, this detailed discussion will lead to an increased awareness and understanding of the model assumptions, 
which is of particular importance, since this aspect may have been somewhat neglected in both applied and theoretical work on the general ANCOVA so far. 
Secondly, we introduce the wild bootstrap method \citep{Wu,Liu,Mam}, combine it with the White-ANCOVA and prove the theoretical validity of this approach.  
Moreover, we present the results of an extensive simulation study, where we investigate the impact
of various degrees of nonnormality and heteroskedasticity on the type I error control
of the ANCOVA F test, the HCCME-based test and its wild bootstrap counterpart in several 
balanced and unbalanced small sample size settings. 
Such scenarios may well be 
encountered in practice, but have not been sufficiently examined in the context of 
ANCOVA so far, not to mention the newly proposed method. Moreover, we simulate the empirical power
of the ANCOVA F test and the wild bootstrap version of the HCCME-based test. Finally, we demonstrate the applicability of 
our method to a real-life dataset and conclude with some discussions of our results, closing remarks and ideas 
for future research. We deliberately put all mathematical proofs into the Online Appendix (Section 1), in order to ensure that the main body of the manuscript can be understood by a broad audience. In Sections  2-6 of the Online Appendix,
we provide additional simulation results for scenarios with alternative HCCMEs as well as for several settings that do not address the main focus of the present manuscript, but nevertheless provide useful empirical evidence regarding the scope of potential applications (\textit{e.g.}, random covariates, or more severe heteroskedasticity). Moreover, the \texttt{R} code for the simulations and the real-data example is also part of the online supplementary material to this paper.       

\section{White's HCCME in the one-way ANCOVA model}\label{SectionWhiteAncova}

In the sequel, we shall denote a diagonal matrix $diag(a_1,a_2,\dots,a_n)$ by the more compact notation $\bigoplus_{i=1}^{n} a_i$, with an analogous notation for a block diagonal matrix (\textit{i.e.}, a matrix consisting of matrices as diagonal elements).  
At first, we introduce the HCCME concept. 
Let us consider the general linear model
\begin{align}
\label{Regression}
{\bf Y} &= \XMatr\boldsymbol{\beta} + \boldsymbol{\epsilon},
\end{align}
where ${\bf Y} = (Y_1,\dots,Y_N)', \boldsymbol{\beta} = (\beta_1,\dots,\beta_c)'$. 
Moreover, let $\XMatr = ({\bf x}_1,\dots,{\bf x}_N)'$  with  ${\bf x}_i = (x_i^{(1)},\dots,x_i^{(c)})'$, $ 1\leq i \leq N$
denote the regressor matrix, considered as fixed in the sequel. The errors are assumed to be independent, with $E(\epsilon_i) = 0$ and $Var(\epsilon_i) = \sigma_i^2 > 0$, $1\leq i \leq N$. Let $\boldsymbol{\hat{\beta}} = (\XMatr'\XMatr)^{-1}\XMatr'{\bf Y}$ denote the ordinary least squares estimator of $\boldsymbol{\beta}$. Moreover, we define
\begin{align}
\label{HCCMEWhite}
\widehat{Cov}(\sqrt{N} \boldsymbol{\hat{\beta}}) &:= (\XMatr'\XMatr/N)^{-1}N^{-1}\XMatr'\bigoplus_{i=1}^{N}u_i^2\XMatr(\XMatr'\XMatr/N)^{-1},
\end{align}
where $u_i^2 := (Y_i - {\bf x}_i^{\prime}\boldsymbol{\hat{\beta}})^2, 1\leq i \leq N.$ 

In order to test $H_0: \MatrFormat{H}\boldsymbol{\beta} = \VecFormat{0}$, where $\MatrFormat{H} \in \mathbb{R}^{q,c}, rank(\MatrFormat{H}) = q,$ White considered a Wald-type test statistic and proved that under certain assumptions, which will be precisely stated in Appendix 1, the following asymptotic result holds \citep{Whi}: 
\begin{align}
\label{TestStatisticWhite}
N\{\MatrFormat{H}(\boldsymbol{\hat{\beta}}-\boldsymbol{\beta})\}'\{\MatrFormat{H}\widehat{Cov}(\sqrt{N} \boldsymbol{\hat{\beta}})\MatrFormat{H}'\}^{-1}\MatrFormat{H}(\boldsymbol{\hat{\beta}}-\boldsymbol{\beta}) \overset{d}{\longrightarrow} \chi_q^2.
\end{align}

Hence, an asymptotic level $\alpha$ test can be obtained by rejecting the null hypothesis $H_0: \MatrFormat{H}\boldsymbol{\beta} = \VecFormat{0}$ if and only if the value of the test statistic 
\begin{align}
\label{TestStatisticWhiteH0}
T(\MatrFormat{H}) := N\{\MatrFormat{H}\boldsymbol{\hat{\beta}}\}'\{\MatrFormat{H}\widehat{Cov}(\sqrt{N} \boldsymbol{\hat{\beta}})\MatrFormat{H}'\}^{-1}\MatrFormat{H}\boldsymbol{\hat{\beta}}
\end{align}
exceeds the $1-\alpha$ quantile of the central Chi-square distribution with $q = rank(\MatrFormat{H})$ degrees of freedom. We would like to mention that
in previous papers, it has been recognized that using White's initial estimator, as defined in \refmath{HCCMEWhite}, makes the 
corresponding test statistics liberal \citep{Mac}. Consequently, some refinements of White's initial estimator had been proposed. 
For example, one could replace the squared residuals $u_i^2$ in \refmath{HCCMEWhite} by $u_i^2/(1-p_{ii})$\citep{Mac} or $u_i^2/(1-p_{ii})^{\delta_i}$, $\delta_i:= \min\left(4, p_{ii} / (N^{-1}\sum_{j=1}^{N}p_{jj})\right)$ \citep{Cri}, where $p_{ii}$ denotes the $i$-th diagonal element of the hat matrix $\XMatr(\XMatr'\XMatr)^{-1}\XMatr'$, $1\leq i\leq N$.  
Note that regarding the proofs provided in Appendix 1, this modification does not
matter, because $\lim_{N \to \infty} (1-p_{ii})^2 = 1$ and $\lim_{N \to \infty}(1-p_{ii})^{\delta_i} = 1$, uniformly in $i$, under the assumptions stated in Appendix 1. Throughout this paper, we refer
to White's initial estimator and its modified versions by HC$0$, HC$2$ and HC$4$, respectively \citep{Mac}. 

Next, to turn to the general one-way ANCOVA model, suppose that we have $N = n_1 + n_2 + \dots + n_a$ observations of individuals from $a$ different groups. We assume that these observations are realizations of random variables, say, $Y_{11}, Y_{12},\dots,Y_{a n_a}$, which follow the linear model 
$Y_{ij} = \mu_i + \sum_{k=1}^{r} \nu_{k}z_{ij}^{(k)} + \epsilon_{ij}$, where $\epsilon_{ij}$ are 
independent with $E(\epsilon_{ij}) = 0,$ $Var(\epsilon_{ij}) = \sigma_{ij}^2 > 0$, and $z_{ij}^{(k)}$ are some fixed covariates, 
$i = 1,\dots,a, j = 1,\dots,n_i, k=1,\dots,r$.  Equivalently, in matrix notation,
\begin{align}
\label{ANCOVA}
{\bf Y}&= \left(\bigoplus_{i=1}^{a} {\bf 1}_{n_i}\right)\boldsymbol{\mu}+\MatrFormat{Z}\boldsymbol{\nu}+\boldsymbol{\epsilon},
\end{align}
where ${\bf 1}_{n_i}$ denotes the $n_i$-dimensional vector of ones, ${\bf Y} = (Y_{11},\dots,Y_{an_a})'$, $\MatrFormat{Z} = ({\bf z}_{11},\dots,{\bf z}_{an_a})'$, 
${\bf z}_{ij} = (z_{ij}^{(1)},\dots,z_{ij}^{(r)})', 1\leq i \leq a, 1\leq j \leq n_i$, 
$\boldsymbol{\mu} = (\mu_1,\dots,\mu_a)'$, 
$\boldsymbol{\nu} = (\nu_1,\dots,\nu_r)'$,
$\boldsymbol{\epsilon} = (\epsilon_{11},\dots,\epsilon_{an_a})'$, 
$E(\boldsymbol{\epsilon}) = {\bf 0}$, 
$Cov(\boldsymbol{\epsilon}) = diag(\sigma_{11}^2, \dots, \sigma_{a n_a}^2).$ 
Now, in order to derive a test for   
\begin{align}
\label{Hypothesis}
H_0: \mu_1 = \dots = \mu_a,
\end{align}
we rewrite the ANCOVA model given 
in \refmath{ANCOVA} in the form of the linear model \refmath{Regression} by setting
$\XMatr = (\bigoplus_{i=1}^{a} {\bf 1}_{n_i}, \MatrFormat{Z}), 
\boldsymbol{\beta} = (\boldsymbol{\mu}',\boldsymbol{\nu}')'$ 
and splitting up the indices in \refmath{Regression}, 
in order to account for the grouped structure of the data. Accordingly, in order to express the hypothesis stated in \refmath{Hypothesis} as $H_0: \MatrFormat{H}\boldsymbol{\beta} = {\bf 0}$, we specify $\MatrFormat{H} = ({\bf 1}_{a-1},-\MatrFormat{I}_{a-1},{\bf 0})$ (\textit{i.e.}, $\MatrFormat{H}$ is a $(1\times 3)$ block matrix, containing the $(a-1)-$dimensional vector ${\bf 1}_{a-1}$ of ones, the $(a-1)-$dimensional identity matrix $\MatrFormat{I}_{a-1}$ multiplied by $(-1)$ and the $((a-1)\times r)$ matrix containing only zeroes, respectively, where $r$ denotes the number of covariates). Observe that $\MatrFormat{H}$ has full row rank 
because $rank(\MatrFormat{H}) = a-1$. Furthermore, \refmath{Hypothesis} is indeed equivalent to $H_0: \MatrFormat{H}\boldsymbol{\beta} = \bf{0}$, because the latter simplifies to $H_0: \mu_1-\mu_2 = 0, \ldots, \mu_1-\mu_a = 0$. Now, in the following theorem, we set up some assumptions, which  are required for the White HCCME approach to work. 

\begin{thm}
\label{PropWhiteAncova}
In what follows, let $\MatrFormat{I}_a$, $\MatrFormat{J}_a$ and $\MatrFormat{P}_a$ denote the $a$-dimensional identity matrix, the $a$-dimensional square matrix of 1's and the so-called $a$-dimensional centering matrix (\textit{i.e.}, $\MatrFormat{P}_a = \MatrFormat{I}_a - \frac{1}{a}\MatrFormat{J}_a$), respectively. Let us assume that the following conditions are fulfilled for model \refmath{ANCOVA}:
{\renewcommand\labelenumi{(GA\theenumi)}
\begin{enumerate}
\item The components of the error vector $\VecFormatGreek{\epsilon}$ are independent, with $E(\boldsymbol{\epsilon}) = {\bf 0}$,\\ $Cov(\boldsymbol{\epsilon}) = diag(\sigma_{11}^2, \sigma_{12}^2,\dots,\sigma_{1n_1}^2,\dots,\sigma_{a1}^2,\dots,\sigma_{an_a}^2)>0$,\\
and $\exists d_1>0, \gamma>0: E(|\epsilon_{ij}|^{2+\gamma})\leq d_1$ for all $i \in \{1,2,\dots,a\}, j \in \{1,2,\dots,n_i\}$. \label{AncovaAssumption1}
\item $\exists d_2>0: |z_{ij}^{(k)}| < d_2$ for all $i \in \{1,\dots,a\}, j \in \{1,\dots,n_i\}, k \in \{1,\dots,r\}$. \label{AncovaAssumption2}
\item
\begin{enumerate}
\item $\exists d_3>0, m_0\in \mathbb{N}: \frac{N}{n_i} \leq d_3$ for all $i \in \{1,2,\dots,a\}$ and $N \geq m_0$.
\item $\exists d_4>0, m_1 \in \mathbb{N}: det(N^{-1}\sum_{i=1}^{a} \MatrFormat{Z}_i'\MatrFormat{P}_{n_i}\MatrFormat{Z}_i) = \prod_{s=1}^{r}\lambda_s \geq d_4$ for all $N\geq m_1$, 
where $\MatrFormat{Z}_i$ denotes the regressor matrix of the $i-$th group, $1\leq i \leq a$, and $\lambda_1,\dots,\lambda_r$ are the 
eigenvalues of the matrix $N^{-1}\sum_{i=1}^{a} \MatrFormat{Z}_i^{\prime}\MatrFormat{P}_{n_i}\MatrFormat{Z}_i$.
\end{enumerate} \label{AncovaAssumption3}
\item
\begin{enumerate} 
\item $\exists d_5 >0, m_2 \in \mathbb{N}: d_5 \sum_{j=1}^{n_i}\sigma_{ij}^2 \geq N$ for all $i \in \{1,2,\dots,a\}$ and $N\geq m_2$.
\item Let $\MatrFormat{M}:= N^{-1}\sum_{i=1}^{a}\MatrFormat{Z}_i^{\prime}\{\MatrFormatGreek{\Sigma}_i - (\sum_{j=1}^{n_i}\sigma_{ij}^2)^{-1}{\bf s}_i{\bf s}_i^{\prime}\}\MatrFormat{Z}_i$, where $\MatrFormatGreek{\Sigma}_i = diag(\sigma_{i1}^2,\dots,\sigma_{in_i}^2)$, 
${\bf s}_i = (\sigma_{i1}^2,\dots,\sigma_{in_i}^2)'$, $1\leq i \leq a$. Then, $\exists d_6 >0, m_3 \in \mathbb{N}$: $det(\MatrFormat{M}) = \prod_{s=1}^{r}\tau_s \geq d_6$ for all $N\geq m_3$,
where $\tau_1,\tau_2,\dots,\tau_r$ denote the eigenvalues
of the matrix $\MatrFormat{M}$.
\end{enumerate} \label{AncovaAssumption4}
\end{enumerate}
} 
Then, the convergence result \refmath{TestStatisticWhite} holds.
\end{thm}

The proof of this theorem is provided in Appendix 1. We would like to emphasize that the assumptions \refmathanc{AncovaAssumption1}-\refmathanc{AncovaAssumption4} are very general, because
they either exclude trivial cases or impose only weak assumptions on the covariates and the errors, which are met in virtually any real-life setting. In particular, observe that the error distributions may even vary between subjects. So, all in all, the proposed method is potentially useful 
for a broad range of applications. This will be further illustrated in the following section.

\subsection{Applicability of the White-ANCOVA model to real-life data}\label{AncovaAssumptions}
At first, we would like to explain how the assumptions 
\refmathanc{AncovaAssumption1}-\refmathanc{AncovaAssumption4} can be interpreted. 
Therefore, we shall discuss the assumptions
stated in Theorem \ref{PropWhiteAncova} point by point now.
\begin{enumerate}[({GA}1)]
\item The errors $\epsilon_{ij}$ are required to be independent with expectation 0. 
These assumptions can be checked by standard techniques ({\it e.g.}, residual plots). 
Independence of observations is most likely justifiable, unless, for example, some patients undergo
a certain examination in the same room/at the same time point. 
The moment condition as stated in \refmathanc{AncovaAssumption1} holds, for example, 
for errors from a symmetric distribution, 
but also for any distribution with finite skewness, which is uniformly bounded. Thus, our model is not very restrictive, allowing for a broad range of distributions, 
including $\chi^2$, $t\, (\text{df}>3)$, exponential, $\Gamma$, and other families of distributions. 
Even more generally, the case of subject-specific error distributions is covered, as long as their third moments are uniformly bounded.  
A typical example would be that the distributions of the errors are allowed to vary between groups, not only in terms of variances, but also in terms of the distribution families they belong to.
\item The uniform boundedness condition on the covariates looks rather restrictive at first sight. However, in real-life settings, most variables do actually satisfy this assumption, because their range of values is indeed bounded for obvious reasons (\textit{e.g.}, blood pressure measurements are neither negative nor arbitrarily large).
\item Assumption (a) states that the group sizes $n_i, 1\leq i \leq a,$ essentially grow at the same rate (\textit{i.e.}, are not extremely unbalanced). 
This assumption is usually needed when doing inferential asymptotics. 
Apart from this formal aspect, the condition also makes sense from a practical point of view: If this assumption was not met, we would, at some point, arrive at a severe under-representation of a certain group, and such a case should be excluded. 

To turn to (b), if the covariates had been random, $N^{-1}\sum_{i=1}^{a}\MatrFormat{Z}_i^{\prime}\MatrFormat{P}_{n_i}\MatrFormat{Z}_i$ would have been the pooled empirical covariance matrix of the covariates. Now, assumption (b) states that this matrix must not be singular for sufficiently large $N$. 
If the empirical covariance matrix was not regular for sufficiently large $N$, this would mean that the variance of either at least one of the covariates or a linear combination of several covariates would converge to 0. 
Thus, that covariate (or the linear combination) would have, asymptotically, a point-mass distribution, which does not make sense from a practical point of view. 
So, even in the more general random covariate scenario, it is reasonable to assume (b). When dealing with fixed covariates, a singular ``empirical covariance matrix'' would mean that either the values of one covariate did not differ between subjects or some covariates were linearly dependent, which would clearly render inclusion of that covariate (these covariates) into the analysis useless.
\item Assumption (a) is met if, for example, the $a$ averages of the error variances within the groups, $n_i^{-1}\sum_{j=1}^{n_i}\sigma_{ij}^2,$ do not become too small if $N$ goes to infinity. To see this, observe that under \refmathanc{AncovaAssumption3}(a), we get $N^{-1}\sum_{j=1}^{n_i}\sigma_{ij}^2 = n_i/N(n_i^{-1}\sum_{j=1}^{n_i}\sigma_{ij}^2) \geq d_3^{-1}(n_i^{-1}\sum_{j=1}^{n_i}\sigma_{ij}^2), 1 \leq i\leq a.$
So, if the latter sum can be uniformly bounded from below for sufficiently large sample sizes, 
this would immediately yield that assumption (a) is met. 
Obviously, such an assumption is sensible because if at least in one particular group, 
the average of the error variances in that group would go to 0, 
the regression part of the ANCOVA would not make sense due to the quasi 
deterministic linear relationship in that group. 

To turn to (b), for sake of simplicity, we assume homogeneity of the variances within the groups, that is, $\sigma_{ij}^2 = \sigma_i^2$ for all $j \in \{1,2,\dots,n_i\}$, $1\leq i \leq a$.
Then, the matrix from \refmathanc{AncovaAssumption4}(b) can be rewritten as
\begin{equation*}
M = \frac{1}{N}\sum_{i=1}^{a}\MatrFormat{Z}_i^{\prime}\{\MatrFormatGreek{\Sigma}_i - (\sum_{j=1}^{n_i}\sigma_{ij}^2)^{-1}{\bf s}_i{\bf s}_i^{\prime}\}\MatrFormat{Z}_i = \frac{1}{N}\sum_{i=1}^{a}\sigma_i^2\MatrFormat{Z}_i^{\prime}\MatrFormat{P}_{n_i}\MatrFormat{Z}_i.
\end{equation*}
Therefore, as long as neither all group-specific error variances nor all within-group variances of at least one
covariate or a linear combination of several covariates are too close to $0$, assumption \refmathanc{AncovaAssumption4}(b) is met.
\end{enumerate}

To close this section, we would like to demonstrate that our model covers a broad range of designs frequently
encountered in practice. In order to keep the notation compact, we use the unique projection
matrix $\MatrFormat{T} = \MatrFormat{H}' (\MatrFormat{H}\MatrFormat{H}')^{-1}\MatrFormat{H}$ to formulate hypotheses
about the vector $\VecFormatGreek{\mu}$ of adjusted means. Note that $\MatrFormat{T}\VecFormatGreek{\beta} = \VecFormat{0} \Leftrightarrow \MatrFormat{H}\VecFormatGreek{\beta} = \VecFormat{0}$,
so, basically, the only change which has to be made is to replace $\MatrFormat{H}$ by $\MatrFormat{T}$ in \refmath{TestStatisticWhite} and
take the Moore-Penrose inverse instead of the classical inverse of the covariance matrix. Observe that the asymptotic result \refmath{TestStatisticWhite} still holds, because the corresponding theorems concerning the distribution of quadratic forms are also valid for the Moore-Penrose inverse \citep{Rav}. 
Furthermore, due to the fact that $\MatrFormat{H} = (\MatrFormat{H}_f,\MatrFormat{0})$, where $\MatrFormat{H}_f$ denotes
the hypothesis matrix corresponding to the factorial part of the parameter vector $\VecFormatGreek{\beta} = (\VecFormatGreek{\mu'},\VecFormatGreek{\nu'})'$,
the corresponding projection matrix is a block diagonal matrix of the form $\MatrFormat{T} = diag(\MatrFormat{T}_f, \MatrFormat{0})$. 
Now, we briefly sketch how hypotheses about factor effects can be tested in several practically
important designs. In what follows, let $\MatrFormat{I}_a$, $\MatrFormat{J}_a$ and $\MatrFormat{P}_a$ denote the $a$-dimensional identity matrix, the $a$-dimensional square matrix of 1's and the so-called $a$-dimensional centering matrix (\textit{i.e.}, $\MatrFormat{P}_a = \MatrFormat{I}_a - \frac{1}{a}\MatrFormat{J}_a$), respectively.

\begin{itemize}
\item {\bf One-way layout.} Suppose you have observations of subjects in $a$ groups (\textit{e.g.}, treatment
arms in a clinical trial). The hypothesis \refmath{Hypothesis}, that is, the null hypothesis of no difference between the adjusted means, can be formulated by
setting $\MatrFormat{T}_f = \MatrFormat{P}_a$.\\
\item {\bf Crossed two-way layout with interactions.} Suppose there are two cross-classified
fixed factors $B$ and $C$ with levels $i = 1,\dots,b$ and $j = 1,\dots,c$ (\textit{e.g.}, the levels of 
$B$ could represent different drugs, whereas the levels of $C$ indicate several dosages, which are required to be the same for all drugs). So, the
total number of factor level combinations is $a=bc$, and by splitting up the indices, 
we have $\VecFormatGreek{\mu} = (\mu_{11},\mu_{12},\dots,\mu_{bc})'$. Using an additive notation, $\mu_{ij} = \mu + \nu_i + \tau_j + (\nu\tau)_{ij}$, with the usual side conditions $\sum_{i}\nu_i = \sum_{j}\tau_j = \sum_{i}(\nu\tau)_{ij} = \sum_{j}(\nu\tau)_{ij} = 0$. The hypotheses of no main effects of 
the factors $B$ (\textit{i.e.}, $\nu_i = 0\,\forall\, i)$ and $C$ (\textit{i.e.}, $\tau_j = 0\,\forall\, j$) can be specified by $\MatrFormat{T}_f = \MatrFormat{P}_b\otimes \frac{1}{c}\MatrFormat{J}_c$
and $\MatrFormat{T}_f = \frac{1}{b}\MatrFormat{J}_b \otimes \MatrFormat{P}_c$, respectively. 
The hypothesis of no interaction effect (\textit{i.e.}, $(\nu\tau)_{ij} = 0\,\forall\, i,j$) is given by $\MatrFormat{T}_f = \MatrFormat{P}_b\otimes \MatrFormat{P}_c$. \\
\item {\bf Hierarchically nested two-way design.} By contrast to the design above, assume now
that $C$ is nested under $B$ (\textit{e.g.}, for each of the drugs $i = 1,\dots,b$, there are specific dosages 
$j = 1,\dots,c_i$ being administered). In this design, the vector of adjusted means is $\VecFormatGreek{\mu} = 
(\mu_{11},\dots,\mu_{1c_1},\dots,\mu_{b1},\dots,\mu_{bc_b})'$. In additive notation, we can write $\mu_{ij} = \mu + \nu_i + \tau(\nu)_{j(i)}$, with the side conditions $\sum_{i}\nu_i = \sum_{j}\tau(\nu)_{j(i)} = 0$. Then, the hypothesis of 
no category effect $B$ (\textit{i.e.}, $\nu_i = 0\,\forall\, i$) and sub-category effect $C(B)$ (\textit{i.e.}, $\tau(\nu)_{j(i)} = 0\,\forall\, i,j$) can be formulated via $\MatrFormat{T}_f = \MatrFormat{P}_{b}\otimes \MatrFormat{\tilde{J}}_c$
and $\MatrFormat{T}_f = \MatrFormat{\tilde{P}}_c$, respectively. Thereby, $\MatrFormat{\tilde{J}}_c = diag(\frac{1}{c_1}\MatrFormat{J}_{c_1},\dots,\frac{1}{c_b}\MatrFormat{J}_{c_b})$ and
$\MatrFormat{\tilde{P}}_c = diag(\MatrFormat{P}_{c_1},\dots,\MatrFormat{P}_{c_b})$. 
\end{itemize}  

The generalization to factorial designs with more than two cross-classified or nested factors
works analogously and is, therefore, not discussed here.

\section{The Wild Bootstrap for the White-ANCOVA model}\label{SectionWb}

Especially in small sample size scenarios, the White-ANCOVA test statistic and the corresponding
asymptotic result stated in \refmath{TestStatisticWhite} might not yield satisfactory results in
terms of maintaining the prespecified type I error probability, see our simulation study in Section~\ref{Simulations} below. A resampling procedure such 
as the bootstrap might remedy this problem. In the context of heteroskedastic regression, 
various so-called wild bootstrap methods have been proposed \citep{Wu, Liu, Mam}. 
The key idea of the wild bootstrap is as follows: Let $u_i^2$ denote the $i$-th squared residual of the linear model \refmath{Regression}, $1\leq i \leq N$. 
Furthermore, let $p_{ii}$ denote the $i$-th diagonal element of the hat matrix $\MatrFormat{P_X} = \XMatr(\XMatr'\XMatr)^{-1}\XMatr'$, $1\leq i \leq N$.
Now, we repeatedly draw random samples consisting of $N$ observations $Y_i^{\ast} = \epsilon_i^{\ast}$, 
where $\epsilon_i^{\ast} = u_i (1-p_{ii})^{-1/2} T_i, 1\leq i \leq N$, $(T_i)_{i=1}^{N}$ are i.i.d. and independently generated from the original data, with
$E(T_1)=0$ and $Var(T_1)=1$. Although for generating the $T_i$'s, one may choose any distribution which satisfies
the latter two conditions,  some particular choices have become popular.
In this paper, we use the Rademacher distribution, which is defined by $P(T_1=-1)=P(T_1=1)= 1/2$.

For each vector ${\bf Y}^{\ast}=(Y_1^{\ast},\dots,Y_N^{\ast})'$ of bootstrap observations,
we calculate the bootstrap OLS estimate $\boldsymbol{\hat{\beta}}^{\ast} = (\XMatr'\XMatr)^{-1}\XMatr'{\bf Y}^{\ast}$ and the bootstrap
version of White's covariance matrix estimator \refmath{HCCMEWhite}, that is, 
\begin{equation}
\label{HCCMEWhiteBoot}
\SigmaWild:= (\XMatr'\XMatr/N)^{-1}N^{-1}\XMatr'\bigoplus_{i=1}^{N}u_i^{\ast 2} \XMatr(\XMatr'\XMatr/N)^{-1},
\end{equation}
where $u_i^{\ast 2} := (Y_i^{\ast} - {\bf x}_i^{\prime}\bestwb)^2, 1\leq i \leq N.$
Finally, we calculate the bootstrap analogon of White's Wald-type test statistic \refmath{TestStatisticWhiteH0}, namely
\begin{align}
\label{TestStatisticWhiteBoot}
T^{\ast}(\MatrFormat{H}) := N\{\MatrFormat{H}\bestwb\}'\{\MatrFormat{H}\SigmaWild \MatrFormat{H}'\}^{-1}\MatrFormat{H}\bestwb.
\end{align}

To turn to the White-ANCOVA setting, we rewrite the ANCOVA model \refmath{ANCOVA} as a special case of the linear model \refmath{Regression}, as we have
already outlined in Section \ref{SectionWhiteAncova}. The main idea of any bootstrap procedure is to resemble the process underlying the generation 
of the original data reasonably well. In the following theorem, we state
that given the data, the distribution of the bootstrap test statistic \refmath{TestStatisticWhiteBoot} 
indeed mimics the distribution of the original test statistic \refmath{TestStatisticWhiteH0} under the 
null hypothesis.
\begin{thm}
\label{TheoremWildBoot}
Let us assume that model \refmath{ANCOVA} as well as the assumptions \refmathanc{AncovaAssumption1}-\refmathanc{AncovaAssumption4} stated in Theorem \ref{PropWhiteAncova} hold. Let $P_{H_0}(T(\MatrFormat{H})\leq x)$ denote the unconditional CDF of $T(\MatrFormat{H})$ under $H_0$ and
$P_{\boldsymbol{\beta}}(T^{\ast}(\MatrFormat{H})\leq x|{\bf Y})$ the conditional CDF of $T^{\ast}(\MatrFormat{H})$ if $\boldsymbol{\beta} \in \mathbb{R}^{a+r}$ is the true underlying parameter. Then, the following statements hold for any $\boldsymbol{\beta} \in \mathbb{R}^{a+r}$.
\begin{enumerate}
\item  $
\sup_{x\in \mathbb{R}}\left|P_{\boldsymbol{\beta}}(T^{\ast}(\MatrFormat{H})\leq x|{\bf Y}) - \chi_q^2(-\infty,x]\right|  \overset{P}{\longrightarrow} 0$ in probability, where 
$q = r(\MatrFormat{H}).$ 
\item $\sup_{x\in \mathbb{R}}\left|P_{\boldsymbol{\beta}}(T^{\ast}(\MatrFormat{H})\leq x|{\bf Y}) - P_{H_0}(T(\MatrFormat{H})\leq x) \right| \overset{P}{\longrightarrow} 0$ in probability.
\end{enumerate} 
\end{thm}  
The proof of this theorem is given in Appendix 1. Note that there, we show that in fact, the wild bootstrap test statistic \refmath{TestStatisticWhiteBoot} yields an asymptotically valid test
in any heteroskedastic linear model under very weak assumptions, which are stated in Appendix 1.

\section{Simulation study}\label{Simulations}
In order to evaluate the finite-sample performance of our proposed tests, we conducted an extensive simulation study, using \texttt{R} version 3.3.1 \citep{Rco17}. 
We assessed the maintenance of a pre-specified alpha level of $5\%$. 
Hereby, we considered an ANCOVA model with four groups and small to moderate sample sizes, 
namely $(n_1,n_2,n_3,n_4)$ $\in$ $\{(40,40,40,40),$ $(15,15,15,15),$ $(5,5,5,5),$ $(5,10,20,25),$ $(25,20,10,5)\}$. 
We assumed two fixed covariate vectors ${\bf z}_1, {\bf z}_2$. 
The first one consisted of equally spaced values between $-10$ and $10$. 
For the second vector, the first and the second half of the components were equally spaced in $[0,5]$ and $[-2,-1]$, respectively, sorted in descending order. The regression coefficients corresponding to the two covariates were assumed to be $-0.5$ and $1.5$, respectively. The vector $\boldsymbol{\mu}$ of the group means was set to ${\bf 0}$, in order to represent an instance of the null hypothesis $H_0: \mu_1 = ... = \mu_4$. 

For each of the sample size scenarios from above, the errors were drawn from the standard normal, 
$\chi_5^2$, lognormal, or double exponential distribution. 
If required, these errors were appropriately shifted and/or scaled and subsequently multiplied with the square root of the covariance matrix $\bigoplus_{i=1}^{a}\sigma_{i}^2\MatrFormat{I}_{n_i}$, 
in order to make sure that the variances of the error terms were indeed equal 
to the values specified as follows: 
For the group-wise error variances, 
we considered the homoskedastic case $\sigma_i^2 = 1$ (Scenario $I$) 
as well as the heteroskedastic setting $\sigma_i^2 = i$, $i\in\{1,2,3,4\}$ (Scenario $II$). 
Note that although we derived the White-ANCOVA tests under the more general assumption of 
subject-specific error variances, such a case would hardly be encountered in practice. 
Most reasonable studies are designed such that the residual variances are rather homogeneous within groups. If this is not the case, it is difficult to interpret the results of the ANCOVA meaningfully. Nevertheless, in order to examine the performance of the White-ANCOVA tests in a more general setting, we also simulated a scenario where within the first group, we assumed a variance of one for the subjects $j = 1,2,...,\lfloor n_1/2 \rfloor$ and a variance of two for the remaining ones, respectively. For the other three groups, we set $\sigma_i^2 = i+1$, where $i = 2,3,4$.
This allocation scenario will be referred to as Scenario $III$. 

Finally, the simulated observations were generated according to \refmath{ANCOVA}. 
For each of the 60 scenario combinations, 
we repeated the data generation process $10\,000$ times, which yields a standard error of $0.22\%$ for the type I error level $\alpha = 5 \%$. 
Within each simulation run, we drew $5\,000$ bootstrap samples, according to the procedure 
described in Section \ref{SectionWb}. 

In addition to the White-ANCOVA test statistic and its wild bootstrap version, 
we also considered the classical ANCOVA F test as a third competitor. It should be noted that the HC$4$ covariance matrix estimator (see Section \ref{SectionWhiteAncova}) was used for both the White-ANCOVA test statistic and its wild bootstrap version. However, we also repeated the simulations for all scenarios using HC$0$ and HC$2$, respectively. The results of the latter can be found in Appendix 2. Table \ref{TableTypeI} contains the simulation results for the HC$4$ estimator.  

\begin{sidewaystable}[h!t]
\centering
\caption{Empirical type I error rates (in $\%$) for the ANCOVA F test, the White-ANCOVA test, and
its wild bootstrap version (based on the HC4 estimator). I: $\sigma_1^2 = \ldots = \sigma_4^2 = 1$, II: $\sigma_i^2 = i, i \in \{1,2,3,4\}$, III: $\sigma_{11}^2 = \ldots = \sigma_{1j}^2 = 1$ and $\sigma_{1(j+1)}^2=\ldots = \sigma_{1n_1}^2 = 2$, $j = \lfloor n_1/2 \rfloor$, $\sigma_i^2 = i+1$ for group $i$, $i \in \{2,3,4\}$. ${\bf n_1} = (40,40,40,40)$, ${\bf n_2} = (15,15,15,15)$, ${\bf n_3} = (5,5,5,5)$, ${\bf n_4}= (5,10,20,25)$, ${\bf n_5} = (25,20,10,5)$. The data generation process was repeated $10\,000$ times. For each simulation run, $5\,000$ bootstrap samples were generated.}
\vspace{5mm}
\begin{tabular}{*{14}{r}}
\hline
\hspace{0.5cm} & \hspace{0.5cm} & \multicolumn{3}{c}{\textbf{Standard normal}} & \multicolumn{3}{c}{\textbf{Standard lognormal}} & \multicolumn{3}{c}{\textbf{Double exponential}} & \multicolumn{3}{c}{{\bf Chi squared ($\text{df} = 5$)}}\\
\cline{3-5}\cline{6-8}\cline{9-11}\cline{12-14}
\textbf{Var}& \textbf{N}& \textbf{F test}& \textbf{White}& \textbf{WB}& \textbf{F test}& \textbf{White}& \textbf{WB}& \textbf{F test}& \textbf{White}& \textbf{WB}& \textbf{F test}& \textbf{White}& \textbf{WB}\\
\hline
$I$ & ${\bf n_1}$ & $5.0$ & $6.2$ & $5.0$ & $4.5$ & $4.9$ & $5.1$ & $5.1$ & $6.2$ & $5.1$ & $4.8$ & $5.9$ & $5.0$\\
& ${\bf n_2}$ & $5.3$ & $9.2$ & $5.5$ & $4.9$ & $5.6$ & $3.7$ & $5.1$ & $8.4$ & $5.2$ & $5.0$ & $8.0$ & $4.9$\\
& ${\bf n_3}$ & $4.5$ & $16.6$ & $5.9$ & $4.2$ & $9.9$ & $3.1$ & $4.9$ & $15.9$ & $6.4$ & $5.0$ & $15.7$ & $5.7$\\
& ${\bf n_4}$ & $5.3$ & $8.2$ & $4.8$ & $4.9$ & $5.9$ & $3.8$ & $4.7$ & $7.5$ & $4.7$ & $5.1$ & $8.3$ & $5.1$\\
& ${\bf n_5}$ & $4.8$ & $8.2$ & $5.0$ & $4.8$ & $4.9$ & $3.1$ & $4.9$ & $7.7$ & $4.9$ & $5.2$ & $8.1$ & $4.5$\\
$II$ & ${\bf n_1}$ & $4.5$ & $6.3$ & $5.1$ & $4.5$ & $5.4$ & $5.7$ & $4.9$ & $6.4$ & $5.3$ & $4.4$ & $6.2$ & $5.0$\\
& ${\bf n_2}$ & $4.8$ & $8.9$ & $5.3$ & $4.7$ & $6.2$ & $4.3$ & $4.9$ & $8.4$ & $5.4$ & $4.9$ & $8.4$ & $5.0$\\
& ${\bf n_3}$ & $4.5$ & $16.5$ & $5.8$ & $4.1$ & $10.1$ & $3.4$ & $5.0$ & $16.1$ & $6.3$ & $4.9$ & $15.7$ & $5.5$\\
& ${\bf n_4}$ & $3.2$ & $8.1$ & $4.8$ & $3.2$ & $6.3$ & $4.7$ & $3.0$ & $7.5$ & $4.8$ & $3.3$ & $8.1$ & $5.2$\\
& ${\bf n_5}$ & $10.1$ & $8.1$ & $5.0$ & $8.2$ & $4.8$ & $3.2$ & $10.0$ & $7.9$ & $4.9$ & $10.5$ & $8.2$ & $4.6$\\
$III$ & ${\bf n_1}$ & $4.8$ & $6.3$ & $5.1$ & $4.8$ & $5.4$ & $5.6$ & $5.0$ & $6.4$ & $5.4$ & $4.6$ & $6.1$ & $5.0$\\
& ${\bf n_2}$ & $5.1$ & $9.1$ & $5.4$ & $4.9$ & $6.0$ & $4.3$ & $5.0$ & $8.6$ & $5.3$ & $4.9$ & $8.4$ & $5.0$\\
& ${\bf n_3}$ & $4.7$ & $16.8$ & $5.7$ & $4.2$ & $10.2$ & $3.1$ & $5.1$ & $16.2$ & $6.3$ & $5.0$ & $16.0$ & $5.5$\\
& ${\bf n_4}$ & $3.5$ & $8.1$ & $4.8$ & $3.4$ & $6.1$ & $4.2$ & $3.3$ & $7.4$ & $4.7$ & $3.5$ & $8.0$ & $5.0$\\
& ${\bf n_5}$ & $9.5$ & $8.2$ & $5.1$ & $7.9$ & $4.9$ & $3.1$ & $9.5$ & $7.9$ & $4.9$ & $10.0$ & $8.3$ & $4.5$\\

\hline
\label{TableTypeI}
\end{tabular}  
\end{sidewaystable}

The White-ANCOVA test tended to be less liberal when it was based on the HC$4$ estimator instead of the HC$0$ or the HC$2$ estimator, whereas the performances of the respective bootstrap versions
were similar to each other. Therefore, the following discussion is focused only on the HC$4$-based tests. In balanced group size scenarios, the classical ANCOVA maintained the prespecified $5\%$ level. In the unbalanced settings, the classical ANCOVA was hardly affected by nonnormality. 
However, heteroskedasticity led to either substantially deflated or inflated 
type I error rates, depending on the relation between the variances and the group sizes. 
In case of positive pairing ({\it i.e.}, the smaller groups 
have the smaller variances), 
the ANCOVA F test tended to be conservative, 
whereas negative pairing ({\it i.e.}, the smaller groups have the larger variances) 
made the test liberal, as suggested by conventional wisdom. These effects became more pronounced when the differences between the group variances were increased (see Appendix 4). The asymptotic White-ANCOVA test yielded substantially inflated type I error rates, both in homo- and heteroskedastic settings. Clearly, the wild bootstrap version maintained the prespecified level well in all scenarios, being superior to the ANCOVA F test in terms of type I error control especially in case of heteroskedasticity and unequal group sizes. 
The slight conservatism seen for
lognormal errors might be caused by the underlying method of estimating the covariance matrix,
since for the White-ANCOVA, we also observed lower type I error rates in the lognormal 
case, compared to the other distributions. Indeed, a closer empirical examination revealed that the variances of the White covariance matrix estimator and its wild bootstrap counterpart were larger for lognormal errors than for the other distributions under consideration (for details, see Appendix 6).\\

\begin{figure}[t]
\centering
\includegraphics[width = 14cm, scale = 1]{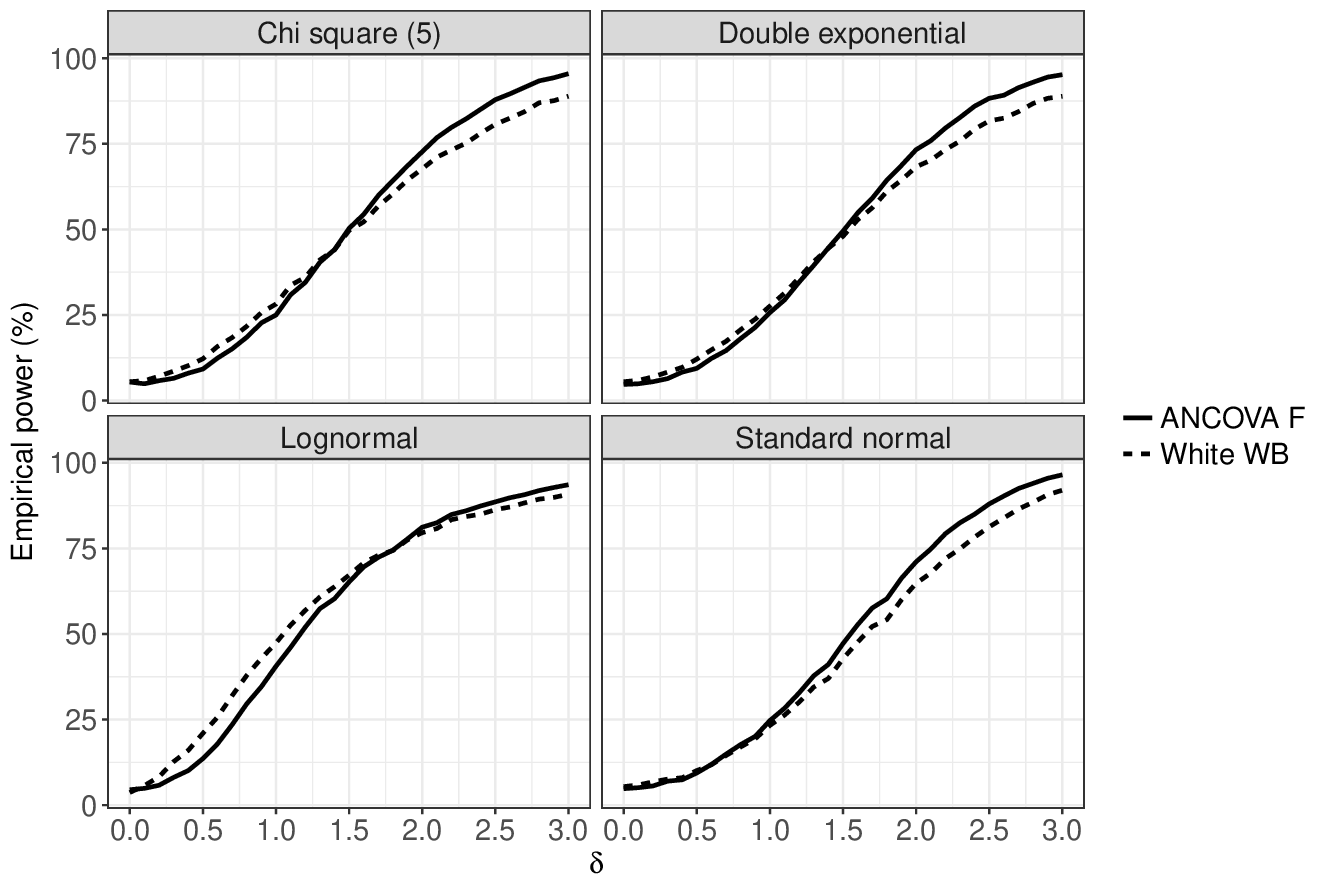}
\caption{Empirical power for the ANCOVA F test (solid) and the wild bootstrap version of the White-ANCOVA test (HC4 version; dashed). Data were generated for two groups with $\mu_1=0$, $\mu_2 = \delta$, $\sigma_1^2 = \sigma_2^2 = 1$, $n_1 = n_2 = 15$.\label{FigurePower}}
\end{figure}

Subsequently, we compared the aforementioned tests with respect to their empirical power. However, 
we only considered the ANCOVA F test and the wild bootstrap version of the White-ANCOVA, 
because the asymptotic White-ANCOVA showed a poor performance in terms of maintaining 
the type I error rates. Furthermore,
in order to make sure that the prespecified level was maintained by both tests, 
we only considered a homoskedastic, balanced setting with $\sigma_1^2 = \sigma_2^2 = 1$ 
and $n_1 = n_2 = 15$. 
Moreover, we specified fixed alternatives by setting $\mu_1 = 0, \mu_2 = \delta$, where
$\delta \in \{0,0.1,\dots,3.0\}$. The four error distributions were chosen as described above. 
For each scenario, we conducted $10\,000$ simulations and $5\,000$ bootstrap runs, respectively. 
The results are displayed in Figure \ref{FigurePower}. 

For small values of $\delta$, 
the wild bootstrap test appeared to be more powerful than the classical ANCOVA. 
As $\delta$ increased, this relationship was gradually being reversed. 
However, the power of the ANCOVA F test at most exceeded
the empirical power of the wild bootstrap test by six to seven 
percentage points. 
So, the bootstrap version of the White-ANCOVA never suffered from a substantial 
power loss compared to the classical ANCOVA test,
even when the assumptions of the latter were met.

Although for ease of presentation, the case of random covariates was not  formally considered in the present manuscript, some simulation results are provided in Appendix 3. All specifications were the same as in the fixed covariate setting, with the only difference that the observations of the second covariate were generated from one of the following distributions: the uniform distribution on $[0,10]$, the standard normal, the standard lognormal, and the $Poisson(\lambda = 5)$ distribution. Overall, the results were very similar to those from the fixed covariate settings. The bootstrap-based method performed even slightly better, being less conservative in case of lognormal errors. Moreover, the maximum power loss of the bootstrap procedure compared to the classical ANCOVA was slightly smaller than in the fixed covariate setting. It should be noted that both with fixed and with random covariates, the somewhat low empirical type I error rates in case of lognormal errors did not translate to an overall loss of power. In fact, for popular choices of the target power (\textit{e.g.}, 80 or 90 percent), the bootstrap-based method had basically the same if not even a slightly better performance (in case of random covariates) compared to the ANCOVA F test (see the lower left panels of Figures \ref{FigurePowerRandUnif}-\ref{FigurePowerRandPoi5} in Section 3 of the Online Appendix).\\

Finally, we also considered some rather extreme scenarios, in order to examine potential limitations of our proposed method. To begin with, especially for the practitioners, it is of interest to give at least some rough advice about sample sizes which are too small for use of the bootstrap-based approach. It is obvious from the simulation results reported in Section 5 of the Online Appendix that our proposed method tended to yield liberal results for total sample sizes of 15 and below. In homoskedastic settings as well as for balanced heteroskedastic scenarios, the ANCOVA F test is recommended. In cases where the group sizes are $5$ and $10$ in groups 1 and 2, respectively, and heteroskedasticity might be present, neither of the three tests stayed close to the target 5 percent level. Secondly, we examined the behavior of the three methods under consideration when the group allocation ratio was more extreme. Again, from Table \ref{TableTypeIseverehet} in the Online Appendix, we clearly see that the effects of positive and negative pairing on the ANCOVA F test were becoming more and more pronounced. The asymptotic White approach and its bootstrap analogue still performed well, with the latter being slightly closer to the target type I error level. Observe, however, that as the group allocation ratio increased (\textit{i.e.}, for $n_1$:$n_2=$ 1:8 or even 1:16), there was some tendency towards either conservative or liberal results, depending on whether positive or negative pairing was present. Still, both White-based approaches outperformed the ANCOVA F test. We would like to mention that from a practical point of view, commonly used allocation ratios are less extreme (\textit{e.g.}, 1:2, 1:3). So, actually, the case of severe group size imbalance might be more interesting from a theoretical rather than from an applied researcher's perspective, unless in cases when, for example, the ANCOVA is used for analyzing data from observational studies, where there might be large differences in subgroup sizes.

\section{Real-life data example}\label{Example}
We illustrate the theoretical considerations from the previous sections by applying the White-ANCOVA as well as its wild bootstrap counterpart to a dataset from a preclinical study of the urological research group and the Institute of Molecular Regenerative Medicine of the Spinal Cord Injury and Tissue Regeneration Center Salzburg (SCI-TReCS), Paracelsus Medical University, Salzburg, Austria. The aim of this study was to assess the efficacy of an anti-inflammatory drug in a rat model of complete spinal cord injury (SCI). After SCI, the bladder is known to turn into a pathological state, as indicated by alterations in cystometric variables, such as the voided volume and detrusor pressure \citep{Mit14}. We would like to emphasize that these pathological changes in human urinary bladder function are among the most serious consequences of SCI, being associated with a substantially decreased quality of life and a high risk of mortality \citep{Cra06, Cru11}. Therefore, the main goal of any pharmacological treatment in SCI is to improve cystometric characteristics towards the pre-SCI level. However, the low prevalence of SCI most likely translates to small sample sizes in studies on SCI patients. Moreover, in preclinical research, it is desirable to sacrifice only a small number of animals, due to ethical reasons. Therefore, in research on spinal cord injury, or more generally, on any rare disease, statistical methods which perform well in small sample sizes are much needed.

We analyse only a subset of the data that was collected. In the sequel, we shall consider a randomized, parallel two-group design, where the rats received either verum (VER, $N = 8$) or placebo (PLAC, $N = 14$). The anti-inflammatory drug or the placebo was administered daily, starting at the day of SCI at a standard dosage. Measurements of the urinary bladder function (besides other parameters the maximum detrusor pressure during voiding in $cm\, H_2O$ and the voided volume in $ml$) were taken prior to SCI (baseline) and at $1,7,14,21$, and $28$ days post injury. For ease of illustration, we shall only consider the maximum detrusor pressure ($P_{det}$) in the sequel. At each time point, $P_{det}$ was obtained as the average over all micturitions within a single time slot of one hour. In the same way, the baseline value for each rat had been determined. 
We consider the difference between the maximum detrusor pressure at $28$ days post
SCI and baseline as outcome and $P_{det}$ at baseline as a covariate,
respectively. This is in line with the recommendations in a recently published EMA guideline \citepalias{Ema15}.

\begin{figure}[t]
\centering
\includegraphics[width=14cm, scale = 1]{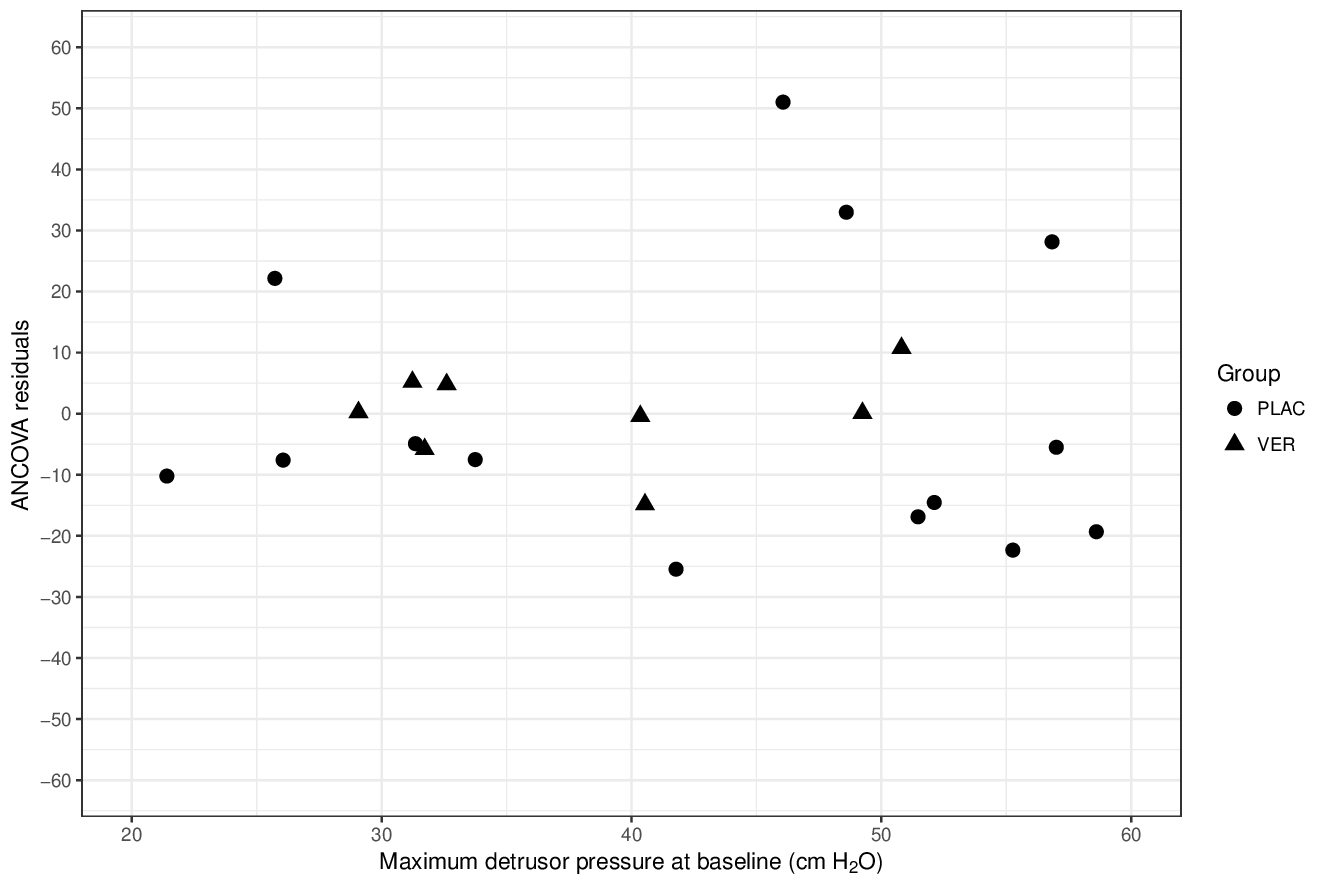}
\caption{Residual plot for the analysis of covariance, with the change in maximum 
detrusor pressure from baseline to 4 weeks post spinal cord injury as outcome and
baseline maximum detrusor pressure as a covariate.}\label{ResidualPlot}
\end{figure}

Before we actually conduct the White-ANCOVA and its wild bootstrap counterpart, we 
take a look at the assumptions \refmathanc{AncovaAssumption1}-\refmathanc{AncovaAssumption4},
which have already been discussed with regard to their interpretation in real-life settings 
in Section \ref{AncovaAssumptions}. To begin with, it is reasonable to assume that the errors have expectation $0$,
because the means of the within-group residuals are close to $0$ ($\bar{u}_{VER}=2.22\times 10^{-16}$,
$\bar{u}_{PLAC} = 3.17\times 10^{-16}$). Moreover, although the residuals for the PLAC group seem to be skewed (see figure \ref{ResidualPlot}), the moment condition stated in \refmathanc{AncovaAssumption1} is most likely met. Note that the skewness itself is not a problem at all, as long as the skewness is uniformly bounded when the sample size grows.  This corresponds to the requirement that the model provides a reasonable fit to the data, which can be ensured by, for example, adding covariates or transformations thereof to the model.     
The independence of the errors, however, cannot be assumed for all subjects, because only one rat each was taken out of ten cages, but two rats each were taken out of six cages.
This issue as well as other potential factors related to the laboratory environment have to
be carefully considered in virtually any preclinical study. However, as this is not the 
main focus of the present work, we shall neglect this aspect for now. Nevertheless, it 
should be emphasized that our proposed method is not capable of accounting for clustered
observations. Therefore, it has to be ensured by properly designing the study that 
clustering is present at most to a negligible extent. 

\begin{figure}[t]
\centering
\includegraphics[width = 14cm, scale = 1]{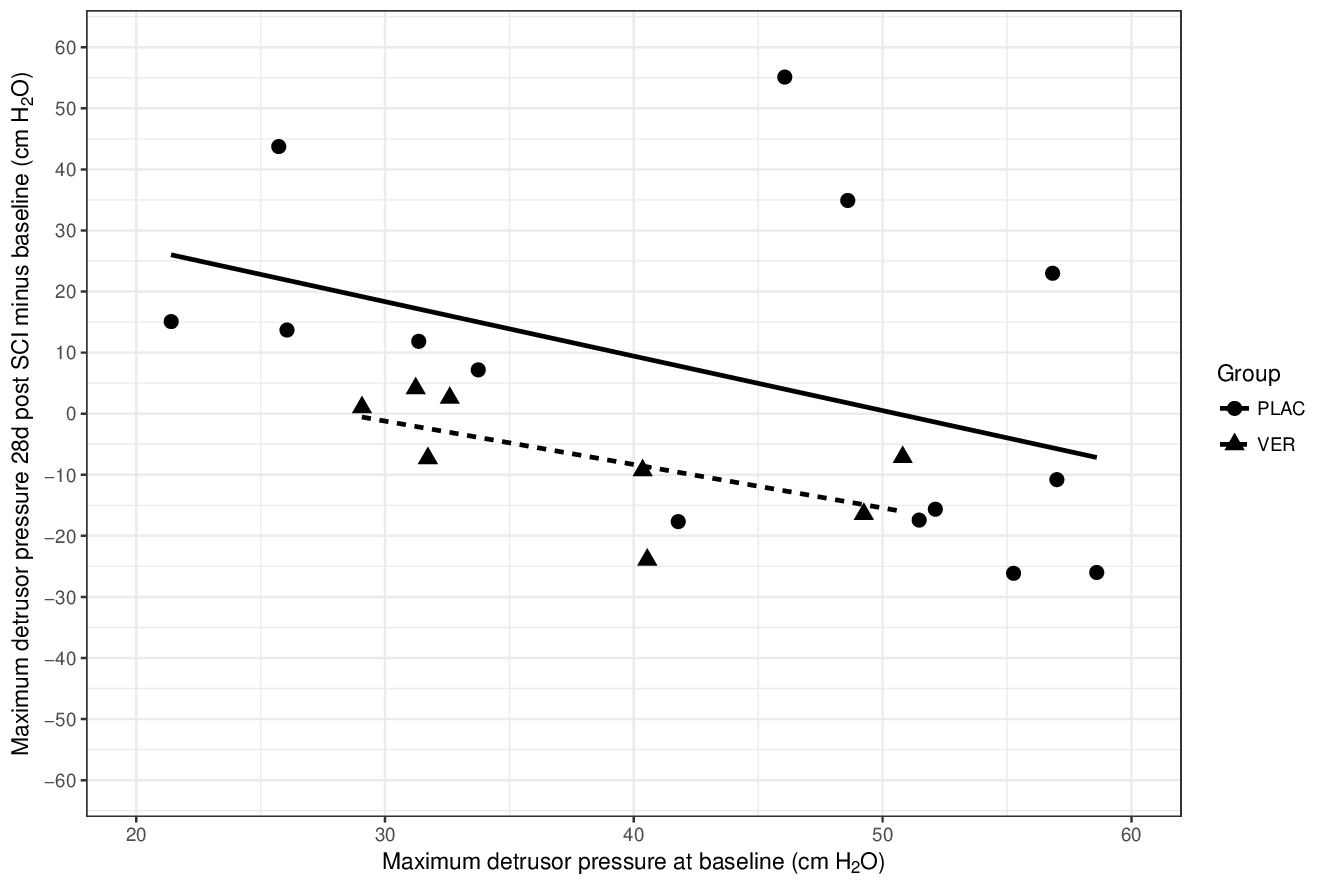}
\caption{Results from the within-group regression of the change of the maximum detrusor pressure from baseline to 4 weeks post 
spinal cord injury on the baseline maximum detrusor pressure.}\label{SlopeComp}
\end{figure}

For physiological reasons, we may rightfully assume that \refmathanc{AncovaAssumption2} holds. 
Assumption \refmathanc{AncovaAssumption3}(a) can be imposed by design, unless 
unforeseen events in the laboratory environment or in the population occur and lead to a 
substantial decrease of observations in one of the groups (\textit{e.g.}, sudden death of a large number
of rats in one group). Furthermore, from figures \ref{ResidualPlot} and \ref{SlopeComp}, it is apparent that 
neither the variance of the baseline maximum detrusor pressure nor the variance of the
residuals are $0$. So, all in all, the assumptions \refmathanc{AncovaAssumption1}-\refmathanc{AncovaAssumption4}
seem justified. 

In general, before conducting an ANCOVA, one has to check if the within-group regression slopes are equal. Indeed, for our data, 
the regression lines are more or less parallel, because $\hat{\beta}_{PLAC} = -0.89\,(SE=0.52)$ and
$\hat{\beta}_{VER} = -0.71\,(SE=0.37)$ (also see figure \ref{SlopeComp}). Therefore, we proceed with the data analysis. Clearly, we have a heteroskedastic 
small sample size setting with positive pairing ($\hat{\sigma}_{VER}^2 = 60.1, \hat{\sigma}_{PLAC}^2 = 558.6$). The ANCOVA F test is not appropriate in such a setting, but the assumptions \refmathanc{AncovaAssumption1}-\refmathanc{AncovaAssumption4} are met. Therefore, we shall compare the results of the White-ANCOVA and its wild bootstrap version now.

Firstly, the estimated adjusted means in the placebo and verum group are $43.69$ and $25.78$, respectively. The White-ANCOVA yields a p value of $0.0078$, whereas the p value resulting from the wild bootstrap White-ANCOVA is $0.0133$. This is in line with the findings of our simulation study, where the asymptotic White-ANCOVA showed a clear tendency towards liberal test decisions. It should be emphasized again that although both tests are asymptotically valid under one and the same set of assumptions \refmathanc{AncovaAssumption1}-\refmathanc{AncovaAssumption4}, they might yield substantially different results in practice, not only in terms of concrete p values, but also with respect to keeping the prespecified type I error rate in small samples, as demonstrated in Section \ref{Simulations}. As this is a serious threat to the validity of the inferential conclusions drawn from the results, caution is apparently needed when applying the White-ANCOVA in heteroskedastic small-sample settings.

\section{Concluding remarks}\label{Conclusions}

As outlined in Section \ref{Intro}, the classical ANCOVA and its bootstrap counterpart as well as the HCCME-based 
approach have been used in many applied research disciplines. 
However, the performance of each of these methods in small samples 
has not been satisfactory, and their combination has not been systematically studied yet. Moreover, in the context of an ANCOVA model, the assumptions underlying the White HCCME approach have neither been thoroughly examined theoretically nor discussed in detail in applied research so far. 
In this paper, we have considered a general ANCOVA model and set up the asymptotic White-ANCOVA test 
statistic as well as its wild bootstrap counterpart. We have proved that under one and the same set of assumptions, both approaches yield asymptotic level $\alpha$ tests.
Note that actually, our proof for the wild bootstrap inference 
does not only cover the ANCOVA, but also the more general case
of a heteroskedastic linear model. In contrast to the work of Mammen, who 
considered an ANOVA-type statistic in a more general case where the model dimension is allowed to vary with
the sample size \citep{Mam}, our proof for the Wald-type statistic uses relatively straightforward techniques. 
Our proposed method relies on rather weak assumptions which are met in virtually any practical situation.
Therefore, it can be utilized in a broad variety of applied research disciplines. Nevertheless, we strongly encourage applied researchers to check the assumptions \refmathanc{AncovaAssumption1}-\refmathanc{AncovaAssumption4} before doing the actual analyses, as illustrated in Section \ref{Example}.

Moreover, the results of the simulations presented in Section \ref{Simulations} indicate that the
direct White-ANCOVA test should not be used in small samples, 
due to severely inflated type I error rates.
However, the wild bootstrap version of the White-ANCOVA showed 
a similar performance as the classical ANCOVA F test in balanced settings 
and outperformed the latter when group sizes
were not equal. The only slight drawback of our proposed test is that it tends to be a bit conservative for errors from a lognormal distribution. However, all in all, we recommend using the wild bootstrap version of the White-ANCOVA
test when group sizes are small and unbalanced. For example, such a situation may well be 
encountered in studies on rare diseases ({\it e.g.}, spinal cord injury) or in preclinical trials. Moreover, our work might also be of considerable relevance for medical centers of small to moderate size. Conducting a trial with a small sample of subjects could be an appealing alternative 
as compared to taking part in a multicenter trial, 
because fewer human and financial resources are needed,
although limited generalizability due to smaller sample sizes could remain as an issue. 

It should be noted that in the context of heteroskedastic regression models, some authors have considered a ``restricted bootstrap'' method, where the residuals are calculated for a model under $H_0$ \citep{Mac10}. In the ANCOVA setting, this would mean that at first, a regression model without group indicators is fitted to the data. The estimates and residuals from that model are used for generating the bootstrap observations, then \citep{Mac10}. We have not considered this approach in the present paper, since our proposed method might be more straightforward and easier to understand for applied researchers. As already mentioned above, the small-sample performance of the ``unrestricted'' approach is very good, so, we think that our proposed method can be recommended for both statistical and practical reasons. 
       
Future research will be aimed at extending the approach presented here in several directions: On the one hand, we will investigate different heteroskedastic multivariate ANCOVA (MANCOVA) methods, particularly focusing on small-sample performance. Moreover, we want to study ANCOVA for clustered data. In this context, the adaption of the cluster-robust covariance matrix estimation techniques proposed by Cameron, Gelbach and Miller \citep{Cam08} together with an examination of its theoretical properties appears to be a promising line of action.

\subsection*{Acknowledgements}
The authors thank Esra Keller and Karin Roider (University Clinic of Urology and Andrology, Spinal Cord Injury and Tissue Regeneration Center, Paracelsus Medical University, Salzburg, Austria) and Ludwig Aigner (Institute of Molecular Regenerative Medicine, Spinal Cord Injury and Tissue Regeneration Center, Paracelsus Medical University, Salzburg, Austria) for generously granting access to their acquired data, which has been used as an illustrative example in Section \ref{Example}. Furthermore, the authors are grateful to the Editor, the Associate Editor and two expert reviewers, for their valuable comments, which improved the quality of the present manuscript considerably.

\subsubsection*{Declaration of conflicting interests}
The authors declared no potential conflicts of interest with respect to the research, authorship, and/or publication of this article.

\subsubsection*{Funding}
The authors disclose receipt of the following financial support for the research, authorship, and/or publication of this article: Arne C. Bathke was supported by the Austrian Science Fund (FWF), grant I 2697-N31 and Markus Pauly by the German Research Foundation (DFG) project DFG-PA 2409/4-1; both within a joined D-A-CH Lead Agency Project.

\subsubsection*{Supplemental material}
\noindent
The following supporting information is available as part of the online article:

\noindent
\textbf{Online Appendix.} This document contains the proofs of the theorems stated in the paper, as well as simulation results for alternative covariance matrix estimators and various additional settings (\textit{e.g.}, more extreme heteroskedasticity, or random covariates). Finally, we also provide some simulations regarding bias and variance of the estimators. 
\\
\noindent
\textbf{Wild\_Bootstrap\_ANCOVA\_Simulations.R.} \texttt{R} code that was used for the simulations presented in Section \ref{Simulations}. 

\noindent
\textbf{Zimmermann\_White\_ANCOVA\_real\_data\_example.R.} \texttt{R} code that was used for the analyses presented in Section \ref{Example}. 

\noindent
\textbf{Zimmermann\_Ancova\_data\_example.csv.} This is the original data that was analysed in Section \ref{Example}.

\clearpage

{
\centering

\Large{{\bf Online Appendix for ``Small-sample performance and underlying assumptions of a bootstrap-based inference method for a general analysis of covariance model with possibly heteroskedastic and nonnormal errors''}}
}
\setcounter{section}{0}

\vspace{1cm}
\noindent
This supplementary material is organized as follows: Appendix 1 contains the proofs of the
mathematical theorems stated in the paper. Appendix 2 provides results of additional simulations using White's initial covariance matrix estimation approach (HC$0$) and an alternative estimator (HC$2$), respectively (see Section \ref{SectionWhiteAncova}). Simulation results for the case of one fixed and one random covariate are presented in Appendix 3. Scenarios with more severe heteroskedasticity and very small or unbalanced group sizes are examined in Appendices 4 and 5, respectively. Finally, Appendix 6 contains some results regarding bias and variance of the proposed estimators.
\newpage

\section{Proofs}

To prove that the asymptotic result stated in (3) indeed holds, 
the following assumptions are required \citep{Whi}.

{\renewcommand\labelenumi{(W\theenumi)}
\begin{enumerate}
\item The components of $\boldsymbol{\epsilon}$ are independent, with $E(\boldsymbol{\epsilon}) = {\bf 0}$, $Cov(\boldsymbol{\epsilon}) = diag(\sigma_1^2, \sigma_2^2,\dots,\sigma_N^2)>0$, and $\exists c_1>0, \gamma>0: E(|\epsilon_i|^{2+\gamma})\leq c_1$ for all $i \in \{1,2,\dots,N\}$.\label{WhiteAssumption1}
\item $\exists c_2>0: |x_i^{(k)}| < c_2$ for all $i \in \{1,\dots,N\}$ and $k\in \{1,\dots,c\}$.\label{WhiteAssumption2}
\item $\exists n_0 \in \mathbb{N}: (N^{-1} \XMatr'\XMatr)^{-1}$ exists and is uniformly bounded element-wise for all $N\geq n_0$.\label{WhiteAssumption3}
\item $\exists n_1 \in \mathbb{N}: (N^{-1} \XMatr' diag(\sigma_1^2,\dots,\sigma_N^2)\XMatr)^{-1}$ exists and is uniformly bounded element-wise for all $N\geq n_1$.\label{WhiteAssumption4}
\end{enumerate}  
} 
In what follows, we will show that \refmathwhite{WhiteAssumption1}-\refmathwhite{WhiteAssumption4} are implied by 
(GA1)-(GA4). Moreover, we shall see that \refmathwhite{WhiteAssumption1}-\refmathwhite{WhiteAssumption4} are also sufficient for the asymptotic validity of the wild bootstrap test statistic, defined in (8). Note that the latter holds true not only for the general ANCOVA, but also for the more general case of a linear model, as defined in (1).

\subsection{Proof of Theorem 1}

\vspace*{0,5cm}

\begin{proof}[Proof of $(GA1)-(GA2) \Rightarrow \refmathwhite{WhiteAssumption1}-\refmathwhite{WhiteAssumption2}$]
This is straightforward to see. 
\end{proof}

\begin{proof}[Proof of $(GA1)-(GA4) \Rightarrow \refmathwhite{WhiteAssumption4}$]
For showing that the assumptions (GA1)-(GA4) are sufficient for \refmathwhite{WhiteAssumption4}, we introduce some notations at first. We partition the matrix $\MatrFormat{Z}$
of the covariates and the covariance matrix $Cov(\boldsymbol{\epsilon})= diag(\sigma_{11}^2,\dots,\sigma_{an_a}^2)$
as follows: $\MatrFormat{Z} = (\MatrFormat{Z}_1^{\prime},\dots,\MatrFormat{Z}_a^{\prime})'$, $Cov(\boldsymbol{\epsilon}) = \bigoplus_{i=1}^{a}\MatrFormatGreek{\Sigma}_i = diag(\MatrFormatGreek{\Sigma}_1, \MatrFormatGreek{\Sigma}_2,\dots,\MatrFormatGreek{\Sigma}_a)$, where $\MatrFormat{Z}_i$ and $\MatrFormatGreek{\Sigma}_i$ denote the matrix of the covariates and of the error variances of the subjects in group $i$,
respectively, $i \in \{1,2,\dots,a\}$.  

We would like
to emphasize that we tried to stay as closely as possible
to White's assumption \refmathwhite{WhiteAssumption4}. Therefore, we explicitely calculate the inverse of $(N^{-1}\XMatr'Cov(\boldsymbol{\epsilon}) \XMatr)$ and simplify the blocks of the resulting matrix. Then, we show that the required conditions indeed hold. 
Recall that in the ANCOVA model, we have 
\begin{align}
\label{BlockMatrix}
\frac{1}{N}\XMatr'Cov(\boldsymbol{\epsilon}) \XMatr &= 
\frac{1}{N}\begin{bmatrix}
			\bigoplus_{i=1}^{a}\left({\bf 1}_{n_i}^{\prime}\MatrFormatGreek{\Sigma}_i{\bf 1}_{n_i}\right) & \left(\bigoplus_{i=1}^{a}{\bf 1}_{n_i}^{\prime}\MatrFormatGreek{\Sigma}_i\right)\MatrFormat{Z} \\
			\MatrFormat{Z}' \left(\bigoplus_{i=1}^{a}\MatrFormatGreek{\Sigma}_i{\bf 1}_{n_i}\right) & \sum_{i=1}^{a}\MatrFormat{Z}_i^{\prime}\MatrFormatGreek{\Sigma}_i \MatrFormat{Z}_i
			\end{bmatrix}.
\end{align}
In order to derive an explicit expression for the inverse of this matrix, we use Schur's formula \citep[p.37, result 2.1.3.2]{Rav}:
\begin{align*}
\left(N^{-1}\XMatr'Cov(\boldsymbol{\epsilon}) \XMatr\right)^{-1} &= 
N \begin{pmatrix}
				\MatrFormat{A} & \MatrFormat{B} \\
				\MatrFormat{C} & \MatrFormat{D}
				\end{pmatrix},
\end{align*}
where
\begin{align*}
\MatrFormat{A}&= \left(\bigoplus_{i=1}^{a}{\bf 1}_{n_i}^{\prime}\MatrFormatGreek{\Sigma_i}{\bf 1}_{n_i}\right)^{-1} + \left(\bigoplus_{i=1}^{a}{\bf 1}_{n_i}^{\prime}\MatrFormatGreek{\Sigma_i}{\bf 1}_{n_i}\right)^{-1}\MatrFormat{BD^{-1}C}\left(\bigoplus_{i=1}^{a}{\bf 1}_{n_i}^{\prime}\MatrFormatGreek{\Sigma_i}{\bf 1}_{n_i}\right)^{-1}\\
\MatrFormat{B}&= - \left(\bigoplus_{i=1}^{a}{\bf 1}_{n_i}^{\prime}\MatrFormatGreek{\Sigma_i}{\bf 1}_{n_i}\right)^{-1}
		\left(\bigoplus_{i=1}^{a}{\bf 1}_{n_i}^{\prime}\MatrFormatGreek{\Sigma_i}\right)\MatrFormat{Z} 
		\MatrFormat{D}, \\
\MatrFormat{C}&= -\MatrFormat{D}
		\MatrFormat{Z}' \left(\bigoplus_{i=1}^{a}\MatrFormatGreek{\Sigma_i}{\bf 1}_{n_i}\right)
		\left(\bigoplus_{i=1}^{a}{\bf 1}_{n_i}^{\prime}\MatrFormatGreek{\Sigma_i}{\bf 1}_{n_i}\right)^{-1},\\ 
D&= \left\{\sum_{i=1}^{a}\MatrFormat{Z}_i^{\prime}\MatrFormatGreek{\Sigma_i} \MatrFormat{Z}_i 
		- \MatrFormat{Z}' 
		\left(\bigoplus_{i=1}^{a}\MatrFormatGreek{\Sigma_i}{\bf 1}_{n_i}\right) 
		\left(\bigoplus_{i=1}^{a}{\bf 1}_{n_i}^{\prime}\MatrFormatGreek{\Sigma_i}{\bf 1}_{n_i}\right)^{-1}
		\left(\bigoplus_{i=1}^{a}{\bf 1}_{n_i}^{\prime}\MatrFormatGreek{\Sigma_i}\right)
		\MatrFormat{Z}
		\right\}^{-1}.
\end{align*}

Now, we simplify $\MatrFormat{A},\MatrFormat{B},\MatrFormat{C}$ and $\MatrFormat{D}$. Let us start with $\MatrFormat{D}$. At first, we show that $N^{-1}\cdot \MatrFormat{D}^{-1}$ is uniformly bounded element-wise under assumptions (GA1) and (GA2). To see this, recall that $\MatrFormat{Z}=(\MatrFormat{Z}_1^{\prime},\dots,\MatrFormat{Z}_a^{\prime})'$ and do some algebra to get
\begin{equation*}
N^{-1}\cdot \MatrFormat{D}^{-1}	= 
				\frac{1}{N}\sum_{i=1}^{a}\MatrFormat{Z}_i^{\prime}\MatrFormatGreek{\Sigma}_i \MatrFormat{Z}_i - 
						\frac{1}{N}\sum_{i=1}^{a}\left(\sum_{j=1}^{n_i}\sigma_{ij}^2\right)^{-1}\MatrFormat{Z}_i^{\prime}{\bf s}_i {\bf s}_i^{\prime}\MatrFormat{Z}_i,
\end{equation*}
where ${\bf s}_i:= (\sigma_{i1}^2,\dots,\sigma_{in_i}^2)'$, $1\leq i \leq a$.
Let $\MatrFormat{Z}_i=({\bf z}_i^{(1)},\dots,{\bf z}_i^{(r)})$, $1\leq i \leq a$.  Applying (GA1) and (GA2) yields 
\begin{equation}
\left|\frac{1}{N} \sum_{i=1}^{a}{\bf z}_i^{(k)\prime}\MatrFormatGreek{\Sigma_i}{\bf z}_i^{(l)}\right| \leq
\frac{1}{N} \sum_{i=1}^{a}\sum_{j=1}^{n_i}\sigma_{ij}^2 \left|z_{ij}^{(k)}\right|\left|z_{ij}^{(l)} \right| \leq d_1 d_2^2
\end{equation}
and
\begin{equation}
\left|\frac{1}{N} \sum_{i=1}^{a} \left(\sum_{j=1}^{n_i}\sigma_{ij}^2\right)^{-1}{\bf z}_i^{(k)\prime} {\bf s}_i {\bf s}_i^{\prime}{\bf z}_i^{(l)}\right| \leq \frac{1}{N}\sum_{i=1}^{a}\left(\sum_{j=1}^{n_i}\sigma_{ij}^2\right)^{-1}\sum_{p=1}^{n_i}\sigma_{ip}^2 \left|z_{ip}^{(k)}\right|\sum_{q=1}^{n_i}\sigma_{iq}^2 \left|z_{iq}^{(l)}\right|
	\leq d_1 d_2^2
\end{equation}
for all $k,l \in \{1,2,\dots,r\}$.
Consequently, $N^{-1}\MatrFormat{D}^{-1}$ is uniformly bounded element-wise.
This implies that $N \cdot \MatrFormat{D} = (N^{-1}\MatrFormat{D}^{-1})^{-1}$ is uniformly bounded element-wise for 
sufficiently large $N$, because $(N^{-1}\MatrFormat{D}^{-1})^{-1} = adj(N^{-1}\MatrFormat{D}^{-1})/det(N^{-1}\MatrFormat{D}^{-1})$, where $adj(N^{-1}\MatrFormat{D}^{-1})$ denotes the adjoint (i.e., the transpose of the cofactor matrix) of $N^{-1}\MatrFormat{D}^{-1}$. The matrix on the right has uniformly bounded elements, because of (GA4)(b) and the fact that $adj(N^{-1}\MatrFormat{D}^{-1})$ is uniformly bounded element-wise (due to the uniform boundedness of the elements of $N^{-1}\MatrFormat{D}^{-1}$). 

Now, let us turn to the block $N\cdot \MatrFormat{C}$. Since $N\cdot \MatrFormat{B} = (N\cdot \MatrFormat{C})'$, it suffices to 
prove that $N\cdot \MatrFormat{B}$ has uniformly bounded elements. Let $\MatrFormat{\tilde{B}}:= \bigoplus_{i=1}^{a}({\bf 1}_{n_i}^{\prime}\MatrFormatGreek{\Sigma_i}{\bf 1}_{n_i})^{-1} \bigoplus_{i=1}^{a}{\bf 1}_{n_i}^{\prime}\MatrFormatGreek{\Sigma_i} \MatrFormat{Z}$. By doing some algebra, we obtain 

\begin{equation*}
\MatrFormat{\tilde{B}} =
	\begin{pmatrix}
		\left(\sum_{p=1}^{n_1}\sigma_{1p}^2\right)^{-1}\sum_{j=1}^{n_1}\sigma_{1j}^2 z_{1j}^{(1)} & \cdots & \left(\sum_{p=1}^{n_1}\sigma_{1p}^2\right)^{-1}\sum_{j=1}^{n_1}\sigma_{1j}^2 z_{1j}^{(r)} \\
		\vdots & \ddots & \vdots \\
		\left(\sum_{p=1}^{n_a}\sigma_{ap}^2\right)^{-1}\sum_{j=1}^{n_a}\sigma_{aj}^2 z_{aj}^{(1)} & \cdots & \left(\sum_{p=1}^{n_a}\sigma_{ap}^2\right)^{-1}\sum_{j=1}^{n_a}\sigma_{aj}^2 z_{aj}^{(r)}  
	\end{pmatrix}.	
\end{equation*}
The elements of the matrix $\MatrFormat{\tilde{B}}$ are uniformly bounded under assumption (GA2). 
This implies that $N\cdot \MatrFormat{B}$ is uniformly bounded element-wise, because $N \cdot \MatrFormat{B} = -\MatrFormat{\tilde{B}}N\MatrFormat{D}$, and both matrices in the latter product are uniformly bounded element-wise and have dimensions independent of $N$.

To complete the proof, we show that $N\cdot \MatrFormat{A}$ is uniformly bounded element-wise: $N\MatrFormat{B}\MatrFormat{D}^{-1}\MatrFormat{C}$ is uniformly bounded element-wise, due to the results we have just shown. 
According to assumption (GA4)(a), the elements of $N(\bigoplus_{i=1}^{a}{\bf 1}_{n_i}^{\prime}\MatrFormatGreek{\Sigma_i}{\bf 1}_{n_i})^{-1}$ are uniformly bounded by a 
constant $d_5$. Thus, we have proved that (GA1)-(GA4) indeed imply \refmathwhite{WhiteAssumption4}. 
\end{proof}

\begin{proof}[Proof of $(GA1)-(GA4)\Rightarrow \refmathwhite{WhiteAssumption3}$]
Due to the fact that the matrix 
$N^{-1}\XMatr'\XMatr$ is just a special case of $N^{-1}\XMatr'Cov(\boldsymbol{\epsilon}) \XMatr$, this statement can be 
proved analogously to above. Therefore, the proof is omitted. However, note that the 
upper-left block of the matrix \refmath{BlockMatrix} simplifies to $\bigoplus_{i=1}^{a}{\bf 1}_{n_i}^{\prime}{\bf 1}_{n_i} = \bigoplus_{i=1}^{a}n_i$. So, when calculating the inverse, we obviously have to make sure that the elements of $diag(N/n_1,\dots,N/n_a)$
are uniformly bounded from above, which is ensured by assumption (GA3)(a). Likewise, (GA3)(b) is required for proving that $\MatrFormat{D}$ has uniformly bounded elements.
\end{proof}

\vspace{1cm}

\subsection{Proof of Theorem 2}
Actually, we prove that the results stated in Theorem 2 hold in the more general setting of a heteroskedastic linear model, assuming that \refmathwhite{WhiteAssumption1}-\refmathwhite{WhiteAssumption4} hold. Applying Theorem 1 yields the validity of Theorem 2, then.\\  Statement 2 in Theorem 2 is implied by Statement 1, because the Chi-square distribution is continuous \citep[p.12, Lemma 2.11]{Van}. So, it is left to show that Statement 1 holds. The proof consists of two main steps:
\begin{enumerate}[{Step }1{:}]
\item Given the data, 
\begin{equation}
\label{WBProofStep1}
N\{\MatrFormat{H}\bestwb\}^{\prime}(\MatrFormat{H}\SigmaFull \MatrFormat{H}^{\prime})^{-1}\MatrFormat{H}\bestwb \overset{d}{\longrightarrow} \chi_{r(\MatrFormat{H})}^2 \quad in\, probability
\end{equation}
where $\SigmaFull:= \{\AvgCovLine\}^{-1}\SigmaPart\{\AvgCovLine\}^{-1}$, $\SigmaPart:= N^{-1} \XMatr' diag(\sigma_1^2,\dots,\sigma_N^2)\XMatr$.
\item $\SigmaWild$ consistently estimates $\SigmaFull$, in the sense that the 
following convergence result holds element-wise:
\begin{equation}
\label{WBProofStep2}
\SigmaWild - \SigmaFull \overset{P}{\longrightarrow} \MatrFormat{0}.
\end{equation}
\end{enumerate}

\begin{proof}[Step 1: Derivation of the asymptotic distribution]
We show that given the data, the expression 
\begin{equation}
\label{WBProof1}
\sqrt{N}\SigmaPart^{-1/2}\AvgCovFrac\bestwb
\end{equation}
is asymptotically multivariate standard normal. To see this, we rewrite \refmath{WBProof1} as follows:
\begin{align}
\label{WBProof2}
\sqrt{N}\SigmaPart^{-1/2}\frac{\XMatr'\XMatr}{N}\bestwb &= 
	\SigmaPart^{-1/2}\frac{1}{\sqrt{N}}\sum_{i=1}^{N}{\bf x}_iu_iT_i/\sqrt{1-p_{ii}}.
\end{align}

Now, in order to show the conditional asymptotic normality of \refmath{WBProof2}, we use a conditional central limit theorem for the wild bootstrap \citep[Theorem A.1]{Bey}. Let ${\bf q}_i:= N^{-1/2}\SigmaPart^{-1/2}{\bf x}_i u_i/\sqrt{1-p_{ii}}$, $1\leq i \leq N$. For the aforementioned theorem to be applied, it suffices to show that 
{\renewcommand\theenumi{\roman{enumi}}
\renewcommand\labelenumi{(\theenumi)}
\begin{enumerate}
\item $\max_{1\leq i \leq N} \left\Vert {\bf q}_i \right\Vert \overset{N \to \infty}{\longrightarrow} 0$ in probability, where $\left\Vert\,\right\Vert$ denotes the Euclidean norm on $\mathbb{R}^{c}$. \label{BootClt01}
\item $\sum_{i=1}^{N}{\bf q}_i{\bf q}_i^{\prime} \overset{N \to \infty}{\longrightarrow} \MatrFormatGreek{\Gamma}$ in probability, where $\MatrFormatGreek{\Gamma}$ denotes some positive definite covariance matrix. \label{BootClt02}
\end{enumerate}
}

To prove \refmath{BootClt01}, it suffices to show that the variances of the residuals $u_i, i \in \{1,2,\dots,N\},$ can be uniformly bounded from above, because $N^{-1/2}$ goes to $0$, and all remaining quantities in ${\bf q}_i$ are uniformly bounded, according to \refmathwhite{WhiteAssumption2}-\refmathwhite{WhiteAssumption4}. 
As our proof uses the very same idea as the proof of Lemma 3 in \citet{Wu}, we just
briefly sketch the main idea here. At first, by using the definition of $u_i$ and some algebra, we find that
\begin{equation}
Var(u_i) = \sigma_i^2(1-p_{ii}) + \sum_{j=1}^{N}p_{ij}^2(\sigma_j^2-\sigma_i^2). \label{BootCltProof01}
\end{equation}
Note that in the last step, we have used the independence of the errors (assumption \refmathwhite{WhiteAssumption1}) and $p_{ii} = \sum_{j=1}^{N}p_{ij}^2$.
Now, if we apply assumption \refmathwhite{WhiteAssumption1} and $p_{ii} = \sum_{j=1}^{N}p_{ij}^2$ again, we immediately
see that the term on the right handside of equation \refmath{BootCltProof01} is indeed uniformly bounded from above. Chebyshev's inequality yields the desired result, then. 


In order to prove \refmath{BootClt02}, we do some algebra to get
\begin{equation}
\sum_{i=1}^{N}{\bf q}_i{\bf q}_i^{\prime} = \SigmaPart^{-1/2}N^{-1}\sum_{i=1}^{N}{\bf x}_i{\bf x}_i^{\prime}u_i^2/(1-p_{ii})\SigmaPart^{-1/2} \\
					= \SigmaPart^{-1/2}\MatrFormat{U}\SigmaPart^{-1/2},
\end{equation}
where $\MatrFormat{U}:= N^{-1}\sum_{i=1}^{N}{\bf x}_i{\bf x}_i^{\prime}u_i^2/(1-p_{ii})$.
Now, we apply a consistency result for $U$ proved by White \citep[Theorem 1]{Whi} and the 
fact that $\lim_{N\to \infty}p_{ii} = 0$. This 
immediately yields that $\sum_{i=1}^{N}{\bf q}_i{\bf q}_i^{\prime} = \SigmaPart^{-1/2}\MatrFormat{U}\SigmaPart^{-1/2} \longrightarrow \MatrFormat{I}_c$ in probability. So, condition \refmath{BootClt02} is fulfilled, too. Therefore, the application of the conditional CLT
for the wild bootstrap \citep[Theorem A.1]{Bey} yields the conditional asymptotic normality of 
\refmath{WBProof2}. Consequently, given the data, the quadratic form 
\begin{equation}
\label{WBProof5}
N\{\MatrFormat{H}\bestwb\}'(\MatrFormat{H}\SigmaFull \MatrFormat{H}')^{-1}\MatrFormat{H}\bestwb
\end{equation}    
has, asymptotically, a central Chi-square distribution with $r(\MatrFormat{H})$ degrees of freedom in probability, where $\SigmaFull = (N^{-1}\XMatr'\XMatr)^{-1}\SigmaPart(N^{-1}\XMatr'\XMatr)^{-1}$. 
\end{proof}
\
\begin{proof}[Step 2: Consistency]


It has already been shown that $\SigmaWhiteFull-\SigmaFull \overset{P}{\longrightarrow} \MatrFormat{0}$ holds \citep[Theorem 1]{Whi}, where
\begin{equation}
\SigmaWhiteFull = \left(\AvgCovFrac\right)^{-1}\frac{1}{N}\XMatr^{\prime}\bigoplus_{i=1}^{N}u_i^2 \XMatr \left(\AvgCovFrac\right)^{-1},
\end{equation}
$u_i:= Y_i-{\bf x}_i^{\prime}\best$, $1\leq i \leq N$.
So, it suffices to show that 
\begin{equation}
\label{WBProof6}
\SigmaWild-\SigmaWhiteFull \overset{P}{\longrightarrow} \MatrFormat{0}. 
\end{equation}
Let us at first recall that
\[
\SigmaWild = \left(\AvgCovFrac\right)^{-1}\frac{1}{N}\XMatr'\bigoplus_{i=1}^{N}u_i^{\ast 2}\XMatr\left(\AvgCovFrac\right)^{-1},
\]
where $\bigoplus_{i=1}^{N}u_i^{\ast 2} = diag(u_1^{\ast 2},u_2^{\ast 2},\dots,u_N^{\ast 2})$ and $u_i^{\ast} = Y_i^{\ast}-{\bf x}_i^{\prime}\bestwb$, $1\leq i \leq N$. 


Because $\{\AvgCovLine\}^{-1}$ is uniformly bounded element-wise due to assumption \refmathwhite{WhiteAssumption3}, \refmath{WBProof6} is implied by  
\begin{equation}
\label{CondConv}
\frac{1}{N}\XMatr^{\prime}\bigoplus_{i=1}^{N}(u_i^{\ast 2} - u_i^{2})\XMatr
	= \frac{1}{N}\sum_{i=1}^{N}(u_i^{\ast 2} - u_i^{2}){\bf x}_i{\bf x}_i^{\prime} \overset{P}{\longrightarrow} {\bf 0}.
\end{equation} 
Note that the unconditional convergence result stated above follows immediately from the conditional version of \refmath{CondConv}, by applying the dominated convergence theorem. 
Therefore, it suffices to show that
for all $s,t \in \{1,2,\dots,c\}$, the following conditions hold:
\begin{enumerate}[(i)]
\item $\lim_{N \to \infty}E\left\{N^{-1}\sum_{i=1}^{N}(u_i^{\ast 2} - u_i^{2})x_{i}^{(s)}x_{i}^{(t)} \condbar \VecFormat{Y}\right\} = 0$ almost surely. \label{ConsistencyCondition01}
\item $\lim_{N \to \infty} Var\left\{N^{-1}\sum_{i=1}^{N}(u_i^{\ast 2} - u_i^{2})x_{i}^{(s)}x_{i}^{(t)} \condbar \VecFormat{Y}\right\} = 0$ almost surely.
\label{ConsistencyCondition02}
\end{enumerate}

In order to prove \refmath{ConsistencyCondition01}, we
at first use $\epsilon_i^{\ast} = u_iT_i/\sqrt{1-p_{ii}}$, $p_{ii} = \VecFormat{x}_i^{\prime}(\mf{X}'\mf{X})^{-1}\VecFormat{x}_i$, as well as  $E(T_i)=0, E(T_iT_j) = 0$ for $i\neq j$ and $Var(T_i)=1$, in order to get 
\begin{align}
E(u_i^{\ast 2}\,|\, \VecFormat{Y}) &= E\left(\left\{\epsilon_i^{\ast} - \VecFormat{x}_i^{\prime}(\mf{X}'\mf{X})^{-1}\mf{X}'\VecFormatGreek{\epsilon^{\ast}}\right\}^2\,|\, \VecFormat{Y}\right)\nonumber\\
&= E(\epsilon_i^{\ast 2}\,|\, \VecFormat{Y}) - 2E\{\epsilon_i^{\ast}{\bf x}_i^{\prime}(\XMatr'\XMatr)^{-1}\XMatr^{\prime}\boldsymbol{\epsilon}^{\ast}\,|\, \VecFormat{Y}\} + E\left\{({\bf x}_i^{\prime}(\XMatr'\XMatr)^{-1}\XMatr^{\prime}\boldsymbol{\epsilon}^{\ast})^2\,|\, \VecFormat{Y}\right\}\nonumber\\
	&= u_i^2/(1-p_{ii}) - 2\sum_{j=1}^{N}\frac{p_{ij}u_iu_j}{\sqrt{(1-p_{ii})(1-p_{jj})}}E\left(T_iT_j\,|\, \VecFormat{Y}\right) + \sum_{j=1}^{N}\frac{p_{ij}^2u_j^2}{1-p_{jj}}\nonumber\\
	&= u_i^2/(1-p_{ii}) - 2p_{ii}\frac{u_i^2}{1-p_{ii}} + \sum_{j=1}^{N}\frac{p_{ij}^2u_j^2}{1-p_{jj}}\nonumber\\
&= u_i^2 - \frac{p_{ii}}{1-p_{ii}}u_i^2 + \sum_{j=1}^{N}\frac{p_{ij}^2u_j^2}{1-p_{jj}}.\label{BootstrapResiduals}	
\end{align}

Now, using \refmath{BootstrapResiduals} yields

\begin{align}
E\left\{\frac{1}{N}\sum_{i=1}^{N}(u_i^{\ast 2} - u_i^{2})x_{i}^{(s)}x_{i}^{(t)} \condbar \VecFormat{Y}\right\} &=
\frac{1}{N}\sum_{i=1}^{N}\left(-\frac{p_{ii}}{1-p_{ii}}u_i^2 + \sum_{j=1}^{N}p_{ij}^2\frac{u_j^2}{1-p_{jj}^2}\right)x_{i}^{(s)}x_{i}^{(t)}. \label{WBProof08}
\end{align}

Observe that $|p_{ij}x_{i}^{(s)}x_{i}^{(t)}| \leq C/N$ for some constant $C>0$, uniformly for $i,j$, holds due to the assumptions \refmathwhite{WhiteAssumption2} and \refmathwhite{WhiteAssumption3}. Moreover, it has been proved in White's paper \citep[Theorem 1]{Whi} that we have
\begin{equation}
\label{WhiteConv}
\frac{1}{N}\sum_{i=1}^{N}(u_i^2-\sigma_i^2) \overset{a.s.}\longrightarrow 0.
\end{equation} 

Since the variances $\sigma_i^2$ are uniformly bounded due to assumption \refmathwhite{WhiteAssumption1}, we thus get
\begin{equation*}
\frac{1}{N^2}\sum_{i=1}^{N}u_i^2 = \frac{1}{N^2}\sum_{i=1}^{N}(u_i^2-\sigma_i^2) + \frac{1}{N^2}\sum_{i=1}^{N}\sigma_i^2 \overset{a.s.}{\longrightarrow} 0.
\end{equation*}

If we apply this to \refmath{WBProof08}, the desired result immediately follows, namely that  
\begin{equation*}
E\left\{\frac{1}{N}\sum_{i=1}^{N}(u_i^{\ast 2} - u_i^{2})x_{i}^{(r)}x_{i}^{(s)} \condbar \VecFormat{Y}\right\} \longrightarrow 0. \quad a.s.,
\end{equation*} 

 This completes the proof of \refmath{ConsistencyCondition01}.\\

To turn to \refmath{ConsistencyCondition02}, it suffices to show that
\begin{equation}
\label{VarSuff}
Var\left\{\frac{1}{N}\sum_{i=1}^{N}(u_i^{\ast 2}-u_i^2) \condbar \VecFormat{Y} \right\} \longrightarrow 0\quad a.s.,
\end{equation}
due to the uniform boundedness assumption \refmathwhite{WhiteAssumption2} on the covariates. 
Obviously, we have
\begin{align*}
Var\left\{\frac{1}{N}\sum_{i=1}^{N}(u_i^{\ast 2}-u_i^2) \condbar \VecFormat{Y} \right\} &= \frac{1}{N^2}Var\left\{\sum_{i=1}^{N}(u_i^{\ast 2}-u_i^2) \condbar \VecFormat{Y} \right\} \\
&= \frac{1}{N^2}\left\{\sum_{i=1}^{N}Var(u_i^{\ast 2} \condbar \VecFormat{Y}) + \sum_{i\neq j} Cov(u_i^{\ast 2},u_j^{\ast 2} \condbar \VecFormat{Y})\right\}.
\end{align*}

Let 
\begin{equation}
\label{AB}
A:= \frac{1}{N^2}\sum_{i=1}^{N}Var(u_i^{\ast 2} \,|\, \VecFormat{Y}),\, B:= \frac{1}{N^2}\sum_{i\neq j} Cov(u_i^{\ast 2},u_j^{\ast 2} \,|\, \VecFormat{Y}).
\end{equation}
Now, we show that $A$ and $B$ both converge to zero almost surely as $N \rightarrow \infty$. Because $\lim_{N\rightarrow \infty}(1-p_{ii}) = 1$, we shall drop the $(1-p_{ii})^{-1/2}$ term for sake of simplicity in the sequel. Therefore, the wild bootstrap error terms simplify to $\epsilon_i^{\ast} = u_i \cdot T_i$, $i = 1,\dots,N$. This implies that 
\begin{equation}
\label{wress}
u_i^{\ast 2} = u_i^2 T_i^2 - 2\sum_{j=1}^{N}u_i u_j T_i T_j p_{ij} + \sum_{k=1}^{N}\sum_{l=1}^{N} u_k u_l T_k T_l p_{ik} p_{il},
\end{equation}

because
\begin{align*}
u_i^{\ast 2} &= \left\{\epsilon_i^{\ast} - \VecFormat{x}_i^{\prime}(\mf{X}'\mf{X})^{-1}\mf{X}'\VecFormatGreek{\epsilon^{\ast}}\right\}^2 \\
&= \epsilon_i^{\ast 2} - 2\epsilon_i^{\ast}\VecFormat{x}_i^{\prime}(\mf{X}'\mf{X})^{-1}\mf{X}'\VecFormatGreek{\epsilon^{\ast}} + \left\{\VecFormat{x}_i^{\prime}(\mf{X}'\mf{X})^{-1}\mf{X}'\VecFormatGreek{\epsilon^{\ast}}\right\}^2\\
&= \epsilon_i^{\ast 2} - 2\sum_{j=1}^{N}\epsilon_i^{\ast}\VecFormat{x}_i^{\prime}(\mf{X}'\mf{X})^{-1}\VecFormat{x}_j\epsilon_j^{\ast} + \left\{\sum_{j=1}^{N}\VecFormat{x}_i^{\prime}(\mf{X}'\mf{X})^{-1}\VecFormat{x}_j\epsilon_j^{\ast}\right\}^2\\
&= u_i^2 T_i^2 - 2\sum_{j=1}^{N}u_i u_j T_i T_j p_{ij} + \sum_{k=1}^{N}\sum_{l=1}^{N} u_k u_l T_k T_l p_{ik} p_{il}.
\end{align*}

At first, we take a look at $A$. Using \refmath{wress} and $T_i^2 = 1$ a.s. yields
\begin{align}
A &= \frac{1}{N^2}\sum_{i=1}^{N}Var(u_i^{\ast 2} \,|\, \VecFormat{Y})\nonumber\\
&= \frac{1}{N^2}\sum_{i=1}^{N}Var\left(u_i^2 T_i^2 - 2\sum_{j=1}^{N}u_i u_j T_i T_j p_{ij} + \sum_{k=1}^{N}\sum_{l=1}^{N} u_k u_l T_k T_l p_{ik} p_{il} \condbar \VecFormat{Y}\right)\nonumber\\
&= \frac{1}{N^2}\sum_{i=1}^{N}Var\left(- 2\sum_{j=1}^{N}u_i u_j T_i T_j p_{ij} + \sum_{k=1}^{N}\sum_{l=1}^{N} u_k u_l T_k T_l p_{ik} p_{il} \condbar \VecFormat{Y}\right)\nonumber\\
&= \frac{1}{N^2} \sum_{i=1}^{N}4\sum_{j=1}^{N}\left\{Var(u_iu_jT_iT_jp_{ij}\,|\,\VecFormat{Y}) + \sum_{m\neq j}Cov(u_iu_jT_iT_jp_{ij},u_iu_mT_iT_mp_{im} \,|\, \VecFormat{Y})\right\}\label{A1}\\
&+ \frac{1}{N^2}\sum_{i,k,l}Var(u_ku_lT_kT_lp_{ik}p_{il}\,|\,\VecFormat{Y})\label{A2} \\
&+ \frac{1}{N^2}\sum_{i=1}^{N}\sum_{(k_1,l_1)\neq(k_2,l_2)}Cov(u_{k_1}u_{l_1}T_{k_1}T_{l_1}p_{ik_1}p_{il_1},u_{k_2}u_{l_2}T_{k_2}T_{l_2}p_{ik_2}p_{il_2} \,|\, \VecFormat{Y})\label{A3}\\
&- \frac{1}{N^2}\sum_{i,j,k,l}2Cov(u_iu_jT_iT_jp_{ij},u_ku_lT_kT_lp_{ik}p_{il}\,|\,\VecFormat{Y}).\label{A4}
\end{align}

Note that we have used Bienayme's equality twice in the last step. Now, due to the fact that $(T_i)_{i\in \mathbb{N}}$ is a sequence of i.i.d. random variables with $E(T_1) = 0$, $Var(T_1) = 1$ and $Var(T_1^2) = 0$, we have 
\begin{equation}
\label{VarRad}
Var(T_s T_t) = 1\, \forall\, s\neq t
\end{equation}

and
\begin{align}
Cov(T_{s_1}T_{t_1},T_{s_2}T_{t_2}) \neq 0 \Leftrightarrow 
\{(s_1 = s_2\,\land\,t_1=t_2)&\,\lor\, (s_1 = t_2\,\land\,s_2 = t_1)\}\,\land\,(s_1\neq t_1).\label{CovRad}
\end{align}

Therefore, \refmath{A1}-\refmath{A4} can be further simplified, as shown in the sequel. Firstly, due to \refmath{CovRad}, we get
\begin{align*}
Cov(u_iu_jT_iT_jp_{ij},u_iu_mT_iT_mp_{im} \,|\, \VecFormat{Y}) = 0\,\forall\, m \neq j,
\end{align*}
and, thus, \refmath{A1} is equal to

\begin{align*}
\frac{4}{N^2}\sum_{i=1}^{N}\sum_{j=1}^{N}Var(u_i u_j T_i T_jp_{ij}\,|\, \VecFormat{Y}) &= \frac{4}{N^2}\sum_{i=1}^{N}\sum_{j=1}^{N}u_i^{2} u_j^2 p_{ij}^{2} Var(T_iT_j\,|\,\VecFormat{Y})\\
& = \frac{4}{N^2}\sum_{i=1}^{N}\sum_{\substack{j=1\\j\neq i}}^{N}u_i^2 u_j^2 p_{ij}^{2},
\end{align*}
where we have used \refmath{CovRad} and the fact that $Var(T_i^2) = 0$ in the last step. \\ 
Next, using the same arguments again, \refmath{A2} can be simplified to
\begin{align*}
\frac{1}{N^2}\sum_{i,k,l}u_k^2u_l^2p_{ik}^2p_{il}^2 Var(T_kT_l\,|\,\VecFormat{Y}) = \frac{1}{N^2}\sum_{\substack{i,k,l\\k\neq l}}u_k^2u_l^2p_{ik}^2p_{il}^2.
\end{align*} 

Thirdly, to turn to \refmath{A3}, \refmath{VarRad} and \refmath{CovRad} yield
\begin{align*}
&\frac{1}{N^2}\sum_{i=1}^{N}\sum_{(k_1,l_1)\neq(k_2,l_2)}Cov(u_{k_1}u_{l_1}T_{k_1}T_{l_1}p_{ik_1}p_{il_1},u_{k_2}u_{l_2}T_{k_2}T_{l_2}p_{ik_2}p_{il_2} \,|\, \VecFormat{Y})\\
& = \frac{1}{N^2}\sum_{i=1}^{N}\sum_{k_1=1}^{N}\sum_{l_1=1}^{N}Var(u_{k_1}u_{l_1}T_{k_1}T_{l_1}p_{ik_1}p_{il_1} \,|\, \VecFormat{Y})\\
& = \frac{1}{N^2}\sum_{i=1}^{N}\sum_{\substack{k,l\\k\neq l}}^{N}u_{k}^2u_l^2p_{ik}^2p_{il}^2.
\end{align*}

Finally, analogous arguments can be applied to simplify \refmath{A4} as follows:
\begin{align*}
&-\frac{1}{N^2}\sum_{i,j,k,l}2Cov(u_iu_jT_iT_jp_{ij},u_ku_lT_kT_lp_{ik}p_{il}\,|\,\VecFormat{Y})\\ = &-\frac{1}{N^2}\sum_{i,j}4Cov(u_iu_jT_iT_jp_{ij},u_iu_jT_iT_jp_{ii}p_{ij}\,|\,\VecFormat{Y})\\
= &-\frac{4}{N^2}\sum_{i\neq j}u_i^2u_j^2p_{ii}p_{ij}^2. 
\end{align*}


All in all, we have derived that
\begin{align}
A &= \frac{1}{N^2}Var(u_i^{\ast 2}\,|\,\VecFormat{Y})\nonumber\\
&= \frac{4}{N^2}\sum_{i\neq j}u_i^2u_j^2p_{ij}^2 + \frac{2}{N^2}\sum_{\substack{i,k,l\\k\neq l}}u_k^2u_l^2p_{ik}^2p_{il}^2 -\frac{4}{N^2}\sum_{i\neq j}u_i^2u_j^2p_{ii}p_{ij}^2.\label{Afinal} 
\end{align}

Now, analogously, we simplify $B$, as defined in \refmath{AB}. To start with, applying \refmath{wress} yields

\begin{align}
B &= \frac{1}{N^2}\sum_{i\neq j}Cov(u_i^{\ast 2},u_j^{\ast 2}\,|\,\VecFormat{Y})\nonumber\\
&=\frac{1}{N^2}\sum_{i\neq j} Cov\left(-2\sum_{l=1}^{N}u_iT_iu_lT_lp_{il},-2\sum_{g=1}^{N}u_jT_ju_gT_gp_{jg}\condbar\VecFormat{Y}\right)\nonumber\\
&-\frac{2}{N^2}\sum_{i\neq j}Cov\left(\sum_{l=1}^{N}u_iT_iu_lT_lp_{il}, \sum_{v,w}u_v u_w T_v T_w p_{jv}p_{jw}\condbar\VecFormat{Y}\right)\nonumber\\
&-\frac{2}{N^2}\sum_{i\neq j} Cov\left(\sum_{m,k}u_mu_kT_mT_kp_{im}p_{ik},\sum_{g=1}^{N}u_jT_ju_gT_gp_{jg}\condbar\VecFormat{Y}\right)\nonumber\\
& + \frac{1}{N^2} \sum_{i\neq j} Cov\left(\sum_{m,k}u_mu_kT_mT_kp_{im}p_{ik},\sum_{v,w}u_v u_w T_v T_w p_{jv}p_{jw}\condbar\VecFormat{Y}\right)\nonumber\\
&= \frac{4}{N^2}\sum_{i\neq j}u_iu_j\sum_{l,g}u_lu_gp_{il}p_{jg}Cov(T_iT_l,T_jT_g\,|\,\VecFormat{Y})\label{B1}\\
&-\frac{2}{N^2}\sum_{i\neq j}u_i\sum_{l,v,w}u_lu_vu_wp_{il}p_{jv}p_{jw}Cov(T_iT_l,T_vT_w\,|\,\VecFormat{Y})\label{B2}\\
&-\frac{2}{N^2}\sum_{i\neq j}u_j\sum_{m,k,g}u_mu_ku_gp_{im}p_{ik}p_{jg}Cov(T_mT_k,T_jT_g\,|\,\VecFormat{Y})\label{B3}\\
&+\frac{1}{N^2}\sum_{i\neq j}\sum_{m,k,v,w}u_mu_ku_vu_wp_{im}p_{ik}p_{jv}p_{jw}Cov(T_mT_k,T_vT_w\,|\,\VecFormat{Y})\label{B4}
\end{align}

Now, we further simplify each of the four parts, by applying \refmath{VarRad} and \refmath{CovRad}. Firstly, \refmath{B1} is thus equal to
\begin{equation*}
\frac{4}{N^2}\sum_{i\neq j}u_iu_ju_ju_ip_{ij}p_{ji}Var(T_iT_j) = \frac{4}{N^2}\sum_{i\neq j}u_i^2u_j^2p_{ij}^2.
\end{equation*}

Secondly, for the very same reasons as provided when simplifying \refmath{A4} before, \refmath{B2} can be simplified to
\begin{equation*}
-\frac{4}{N^2}\sum_{i\neq j} u_i\sum_{w=1}^{N}u_w^2u_ip_{iw}p_{ji}p_{jw} Cov(T_iT_w,T_iT_w)= -\frac{4}{N^2}\sum_{i\neq j} u_i^2\sum_{\substack{w=1\\w\neq i}}^{N}u_w^2p_{iw}p_{ij}p_{jw}.
\end{equation*}

Analogously, it can be derived that \refmath{B3} is equal to 
\begin{equation*}
-\frac{4}{N^2}\sum_{i\neq j}u_j^2\sum_{m=1}^{N}u_m^2p_{im}p_{ij}p_{jm}.
\end{equation*}

Likewise, it turns out that \refmath{B4} can be simplified to 
\begin{equation*}
\frac{2}{N^2}\sum_{i\neq j}\sum_{m\neq k}u_m^2u_k^2p_{im}p_{ik}p_{jm}p_{jk}.
\end{equation*}

To sum things up, we thus get
\begin{align*}
B &= Cov(u_i^{\ast 2},u_j^{\ast 2}\,|\,\VecFormat{Y}) \\
&=  \frac{4}{N^2}\sum_{i\neq j}u_i^2u_j^2p_{ij}^2-\frac{4}{N^2}\sum_{i\neq j} u_i^2\sum_{\substack{w=1\\w\neq i}}^{N}u_w^2p_{iw}p_{ij}p_{jw}\\ 
&-\frac{4}{N^2}\sum_{i\neq j}u_j^2\sum_{m=1}^{N}u_m^2p_{im}p_{ij}p_{jm} + \frac{2}{N^2}\sum_{i\neq j}\sum_{m\neq k}u_m^2u_k^2p_{im}p_{ik}p_{jm}p_{jk}.
\end{align*}
  
Consequently, considering \refmath{AB} and \refmath{Afinal}, we have
\begin{align*}
Var\left(\frac{1}{N}\sum_{i=1}^{N}(u_i^{\ast 2}-u_i^2) \condbar \VecFormat{Y} \right)&= A + B \\
&= \frac{4}{N^2}\sum_{i\neq j}u_i^2u_j^2p_{ij}^2 + \frac{2}{N^2}\sum_{\substack{i,k,l\\k\neq l}}u_k^2u_l^2p_{ik}^2p_{il}^2 \\
&-\frac{4}{N^2}\sum_{i\neq j}u_i^2u_j^2p_{ii}p_{ij}^2 + \frac{4}{N^2}\sum_{i\neq j}u_i^2u_j^2p_{ij}^2\\
&-\frac{4}{N^2}\sum_{i\neq j} u_i^2\sum_{\substack{w=1\\w\neq i}}^{N}u_w^2p_{iw}p_{ij}p_{jw}-\frac{4}{N^2}\sum_{i\neq j}u_j^2\sum_{m=1}^{N}u_m^2p_{im}p_{ij}p_{jm}\\
& + \frac{2}{N^2}\sum_{i\neq j}\sum_{m\neq k}u_m^2u_k^2p_{im}p_{ik}p_{jm}p_{jk}.
\end{align*}

Now, if we can show that each sum converges to $0$ a.s., we are done. Firstly, note that due to the assumptions \refmathwhite{WhiteAssumption2} and \refmathwhite{WhiteAssumption3}, $|p_{ij}| \leq C/N$ for some $C>0$, for all $i,j \in \{1,2,\dots,N\}$. Together with assumption \refmathwhite{WhiteAssumption1}, this immediately yields the desired result. As the respective proofs work analogously, we provide details only for one of the sums from above. At first, we do some algebra to get
\begin{align}
\frac{2}{N^2}\sum_{i\neq j}\sum_{m\neq k}u_m^2u_k^2|p_{im}||p_{ik}||p_{jm}||p_{jk}|&\leq \frac{2}{N^6}\sum_{i,j}\sum_{k,m}u_m^2u_k^2\nonumber\\
&=\frac{2}{N^4}\sum_{k,m}u_m^2u_k^2\nonumber\\
&= 2\left(\frac{1}{N^2}\sum_{k=1}^{N}u_k^2\right)\left(\frac{1}{N^2}\sum_{m=1}^{N}u_m^2\right).\label{SumConv}
\end{align} 

According to Theorem 1 in \citet{Whi}, we have 
\begin{equation*}
\frac{1}{N}\left(\sum_{k=1}^{N}u_i^2 - \sum_{k=1}^{N}\sigma_k^2\right) \overset{N\to \infty}{\longrightarrow} 0\quad a.s.
\end{equation*} 

Moreover, assumption \refmathwhite{WhiteAssumption1} yields
\begin{equation*}
\frac{1}{N^2}\sum_{k=1}^{N}\sigma_k^2 \overset{N\to \infty}{\longrightarrow} 0.
\end{equation*} 

Consequently, the expression given in \refmath{SumConv} converges to $0$ almost surely as $N$ goes to infinity. Analogously, it can be proved that the remaining parts of the additive decomposition of $Var(u_i^{\ast 2}-u_i^2|\VecFormat{Y})$ displayed above converge to $0$ almost surely.

Summing up, we have shown that the conditions \refmath{ConsistencyCondition01} and \refmath{ConsistencyCondition02} both hold. Consequently, we get
\[
\SigmaWild-\SigmaWhiteFull \overset{P}{\longrightarrow} \MatrFormat{0}.
\]

Finally, to put everything together, we apply the subsequence principle for convergence in probability. For every sequence of indices $(n_k)$, we can find a subsequence $(n_{k_l})$ such that \refmath{WBProofStep1} holds almost surely along this subsequence. Using  \refmath{WBProofStep2} and Slutzky's theorem, we thus get that conditional on the data,
\[
N\{\MatrFormat{H}\bestwb\}^{\prime}(\MatrFormat{H}\SigmaWild \MatrFormat{H}^{\prime})^{-1}\MatrFormat{H}\bestwb \overset{d}{\longrightarrow} \chi_{r(\MatrFormat{H})}^2 \quad a.s.
\]
along the sequence $(n_{k_l})$. Since $(n_k)$ was chosen arbitrarily, applying the subsequence principle completes the proof of Statement 1 in Theorem 2.
\end{proof}

\newpage

\section{Further simulation results: Alternative heteroskedasticity-consistent covariance matrix estimators}

\vspace{1cm}

As mentioned in Section \ref{Simulations}, all scenarios were simulated three times, one time using the HC$0$ estimator, the other time using the HC$2$ estimator, and finally using the HC$4$ estimator of the covariance matrix. The results of the former two approaches are reported in Table \ref{TableTypeIHC0} and Table \ref{TableTypeIHC2}, respectively; the results based on the HC$4$ estimator are presented in the manuscript. The simulation setup was the same for all three variants (for details, see Section \ref{Simulations}). 

\begin{table}[h!b]
\caption{Empirical type I error rates (in $\%$) for the White-ANCOVA test (White) and its
wild bootstrap version (WB), based on the HC$0$ covariance matrix estimator.}
\vspace{5mm}
\centering
\begin{tabular}{*{10}{r}}
\hline
\hspace{0.2cm} & \hspace{0.2cm} & \multicolumn{2}{c}{\textbf{Normal}} & \multicolumn{2}{c}{\textbf{Lognormal}} & \multicolumn{2}{c}{\textbf{Double exp.}} & \multicolumn{2}{c}{\textbf{Chi square($5$)}}\\
\cline{3-4}\cline{5-6}\cline{7-8}\cline{9-10}
{\bf Var} & {\bf N} & {\bf White} & {\bf WB} & {\bf White} & {\bf WB} & {\bf White} & {\bf WB} & {\bf White} & {\bf WB}\\
\hline
$I$ & ${\bf n_1}$ & $7.2$ & $5.2$ & $5.5$ & $4.7$ & $7.3$ & $5.2$ & $7.1$ & $5.1$ \\
	& ${\bf n_2}$ & $11.9$ & $5.1$ & $8.4$ & $3.4$ & $10.8$ & $5.0$ & $11.5$ & $5.0$ \\
	& ${\bf n_3}$ & $31.9$ & $6.9$ & $23.5$ & $3.5$ & $30.9$ & $6.1$ & $31.1$ & $6.0$ \\
	& ${\bf n_4}$ & $12.8$ & $5.1$ & $9.8$ & $3.6$ & $12.1$ & $4.9$ & $12.4$ & $4.8$ \\
	& ${\bf n_5}$ & $13.5$ & $5.0$ & $8.9$ & $3.1$ & $13.4$ & $4.7$ & $11.6$ & $4.3$ \\
$II$& ${\bf n_1}$ & $7.2$ & $5.1$ & $6.2$ & $5.4$ & $7.2$ & $5.3$ & $7.1$ & $4.9$ \\
	& ${\bf n_2}$ & $12.1$ & $5.4$ & $9.1$ & $3.9$ & $11.3$ & $5.0$ & $12.2$ & $4.9$ \\
	& ${\bf n_3}$ & $32.5$ & $7.0$ & $24.3$ & $3.8$ & $30.7$ & $6.3$ & $31.3$ & $5.9$ \\
	& ${\bf n_4}$ & $12.7$ & $5.1$ & $10.6$ & $4.6$ & $11.8$ & $5.0$ & $12.1$ & $4.8$ \\
	& ${\bf n_5}$ & $13.8$ & $5.1$ & $9.4$ & $3.2$ & $13.5$ & $4.8$ & $12.0$ & $4.4$ \\
$III$& ${\bf n_1}$ & $7.4$ & $5.1$ & $6.2$ & $5.4$ & $7.3$ & $5.4$ & $7.2$ & $5.1$ \\
	& ${\bf n_2}$ & $12.1$ & $5.3$ & $9.4$ & $3.8$ & $11.3$ & $4.9$ & $11.9$ & $4.9$ \\
	& ${\bf n_3}$ & $32.8$ & $7.1$ & $24.0$ & $3.6$ & $30.9$ & $6.2$ & $31.8$ & $5.8$ \\
	& ${\bf n_4}$ & $12.6$ & $5.1$ & $10.4$ & $4.5$ & $11.9$ & $5.2$ & $12.2$ & $4.8$ \\
	& ${\bf n_5}$ & $14.0$ & $4.9$ & $9.3$ & $3.1$ & $13.7$ & $4.8$ & $12.2$ & $4.5$ \\
\hline
\end{tabular}
\label{TableTypeIHC0}
\end{table}

\begin{table}[h!b]
\centering
\caption{Empirical type I error rates (in $\%$) for the White-ANCOVA test (White) and
its wild bootstrap version (WB), based on the HC$2$ covariance matrix estimator.}
\vspace{5mm}
\begin{tabular}{*{10}{r}}
\hline
\hspace{0.2cm} & \hspace{0.2cm} & \multicolumn{2}{c}{\textbf{Normal}} & \multicolumn{2}{c}{\textbf{Lognormal}} & \multicolumn{2}{c}{\textbf{Double exp.}} & \multicolumn{2}{c}{\textbf{Chi square($5$)}}\\
\cline{3-4}\cline{5-6}\cline{7-8}\cline{9-10}
{\bf Var} & {\bf N} & {\bf White} & {\bf WB} & {\bf White} & {\bf WB} & {\bf White} & {\bf WB} & {\bf White} & {\bf WB}\\
\hline
$I$ & ${\bf n}_1$ & $6.4$ & $5.2$ & $4.9$ & $4.8$ & $6.4$ & $5.1$& $6.2$ & $5.1$ \\
	& ${\bf n}_2$ & $8.9$ & $5.1$ & $5.9$ & $3.5$ & $8.2$ & $5.2$ & $8.9$ & $5.1$ \\
	& ${\bf n}_3$ & $19.6$ & $6.9$ & $12.6$ & $3.5$ & $18.1$ & $6.2$ & $17.4$ & $6.1$ \\
	& ${\bf n}_4$ & $9.6$ & $5.2$ & $6.8$ & $3.6$ & $8.8$ & $4.8$& $9.0$ & $4.8$ \\
	& ${\bf n}_5$ & $9.6$ & $5.1$ & $5.8$ & $3.0$ & $9.6$ & $4.8$ & $8.4$ & $4.2$ \\
$II$& ${\bf n}_1$ & $6.3$ & $5.1$ & $5.5$ & $5.4$ & $6.6$ & $5.2$& $6.2$ & $5.1$ \\
	& ${\bf n}_2$ & $8.9$ & $5.3$ & $6.4$ & $3.9$ & $8.3$ & $5.0$& $9.1$ & $5.3$ \\
	& ${\bf n}_3$ & $19.4$ & $6.9$ & $13.6$ & $3.7$ & $18.2$ & $6.2$ & $18.4$ & $6.3$ \\
	& ${\bf n}_4$ & $9.3$ & $5.2$ & $7.5$ & $4.6$ & $8.8$ & $5.0$& $8.7$ & $4.9$ \\
	& ${\bf n}_5$ & $9.8$ & $5.0$ & $6.0$ & $3.1$ & $9.9$ & $4.9$ & $9.6$ & $5.0$ \\
$III$& ${\bf n}_1$ & $6.4$ & $5.1$ & $5.4$ & $5.4$ & $6.5$ & $5.4$& $6.2$ & $5.1$ \\
	& ${\bf n}_2$ & $9.0$ & $5.3$ & $6.3$ & $3.7$ & $8.3$ & $4.9$& $8.8$ & $5.0$ \\
	& ${\bf n}_3$ & $19.6$ & $7.1$ & $13.1$ & $3.6$ & $18.2$ & $6.1$ & $18.1$ & $5.7$ \\
	& ${\bf n}_4$ & $9.4$ & $5.1$ & $7.2$ & $4.5$ &$8.5$ & $5.2$& $9.2$ & $4.8$ \\
	& ${\bf n}_5$ & $9.7$ & $4.9$ & $5.9$ & $3.1$ & $9.6$ & $4.9$ & $8.3$ & $4.4$ \\
\hline
\end{tabular}  
\label{TableTypeIHC2}
\end{table}

\clearpage

\section{Further simulation results: One fixed and one random covariate}

\vspace{1cm}

\begin{figure}[t]
\centering

\includegraphics[width = 16cm, scale = 1]{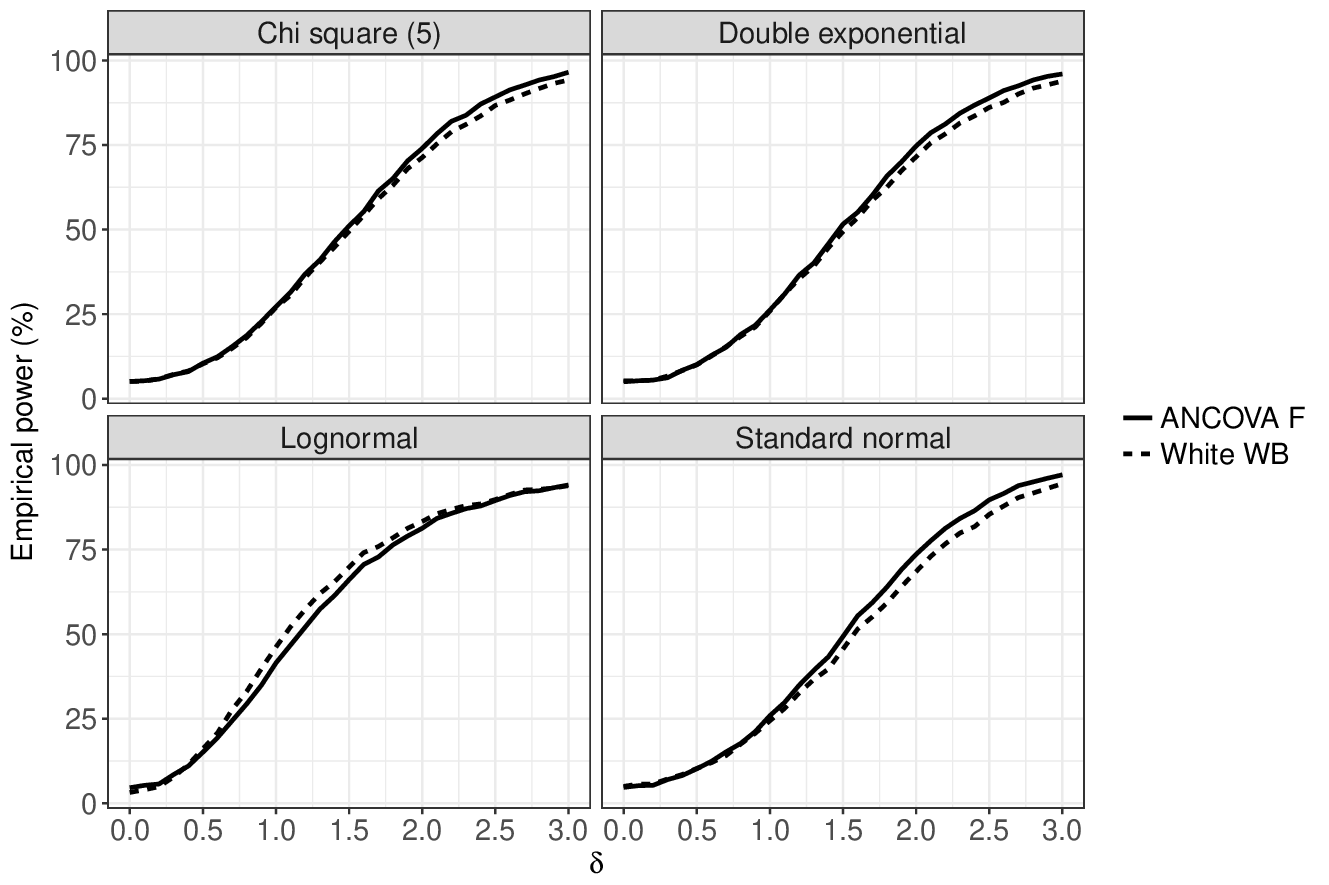}
\caption{Empirical power for the ANCOVA F test (solid) and the wild bootstrap version of the White-ANCOVA test (HC4 version; dashed). Data were generated for two groups with $\mu_1=0$, $\mu_2 = \delta$, $\sigma_1^2 = \sigma_2^2 = 1$, $n_1 = n_2 = 15$. The values of the first covariate were equally spaced between $-10$ and $10$, whereas the values of the second covariate were generated from a uniform distribution on $[0,10]$.\label{FigurePowerRandUnif}}
\end{figure}

\begin{figure}[t]
\centering
\includegraphics[width = 16cm, scale = 1]{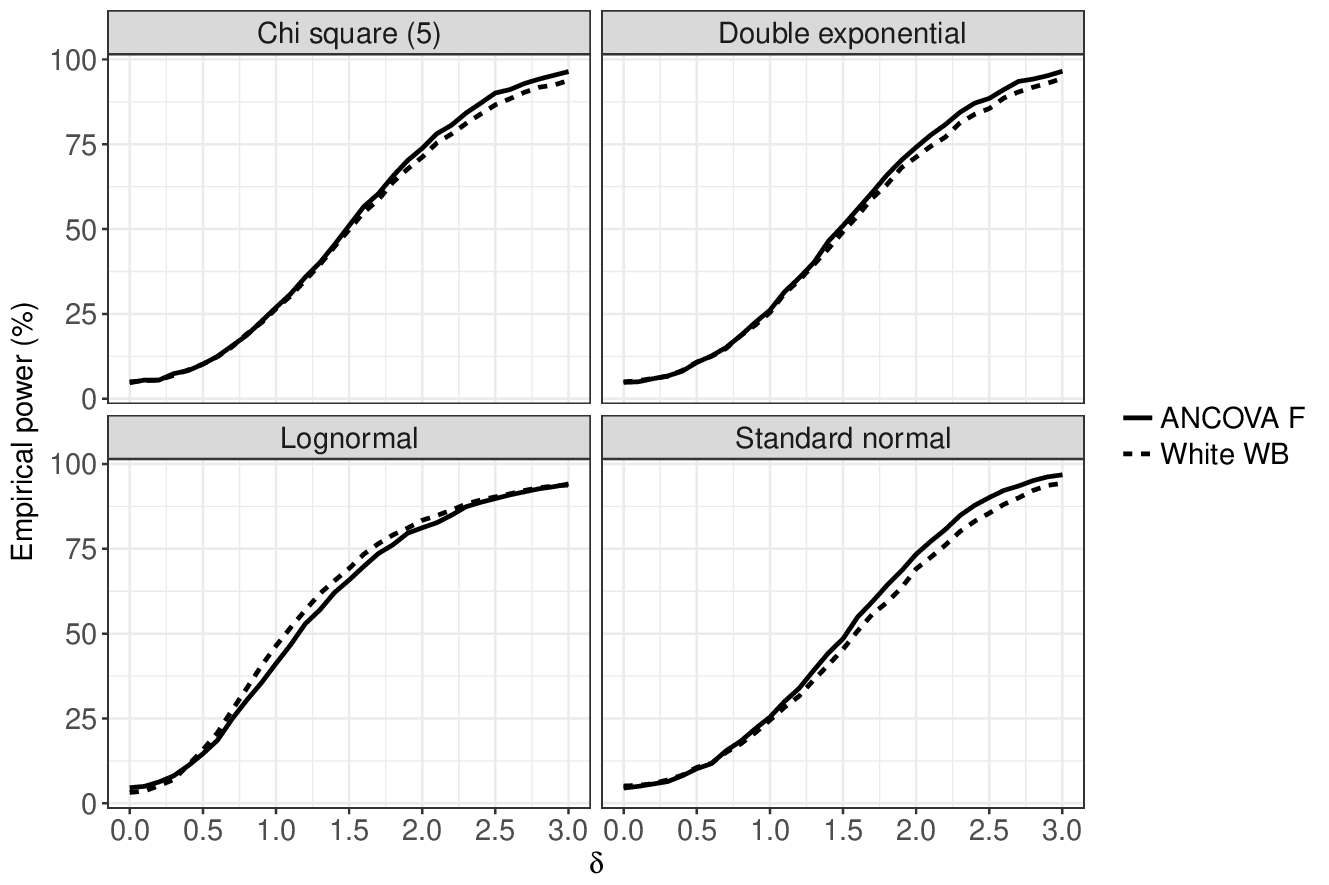}
\caption{Empirical power for the ANCOVA F test (solid) and the wild bootstrap version of the White-ANCOVA test (HC4 version; dashed). Data were generated for two groups with $\mu_1=0$, $\mu_2 = \delta$, $\sigma_1^2 = \sigma_2^2 = 1$, $n_1 = n_2 = 15$. The values of the first covariate were equally spaced between $-10$ and $10$, whereas the values of the second covariate were generated from a standard normal distribution.\label{FigurePowerRandNormal}}
\end{figure}

\begin{figure}[t]
\centering
\includegraphics[width = 16cm, scale = 1]{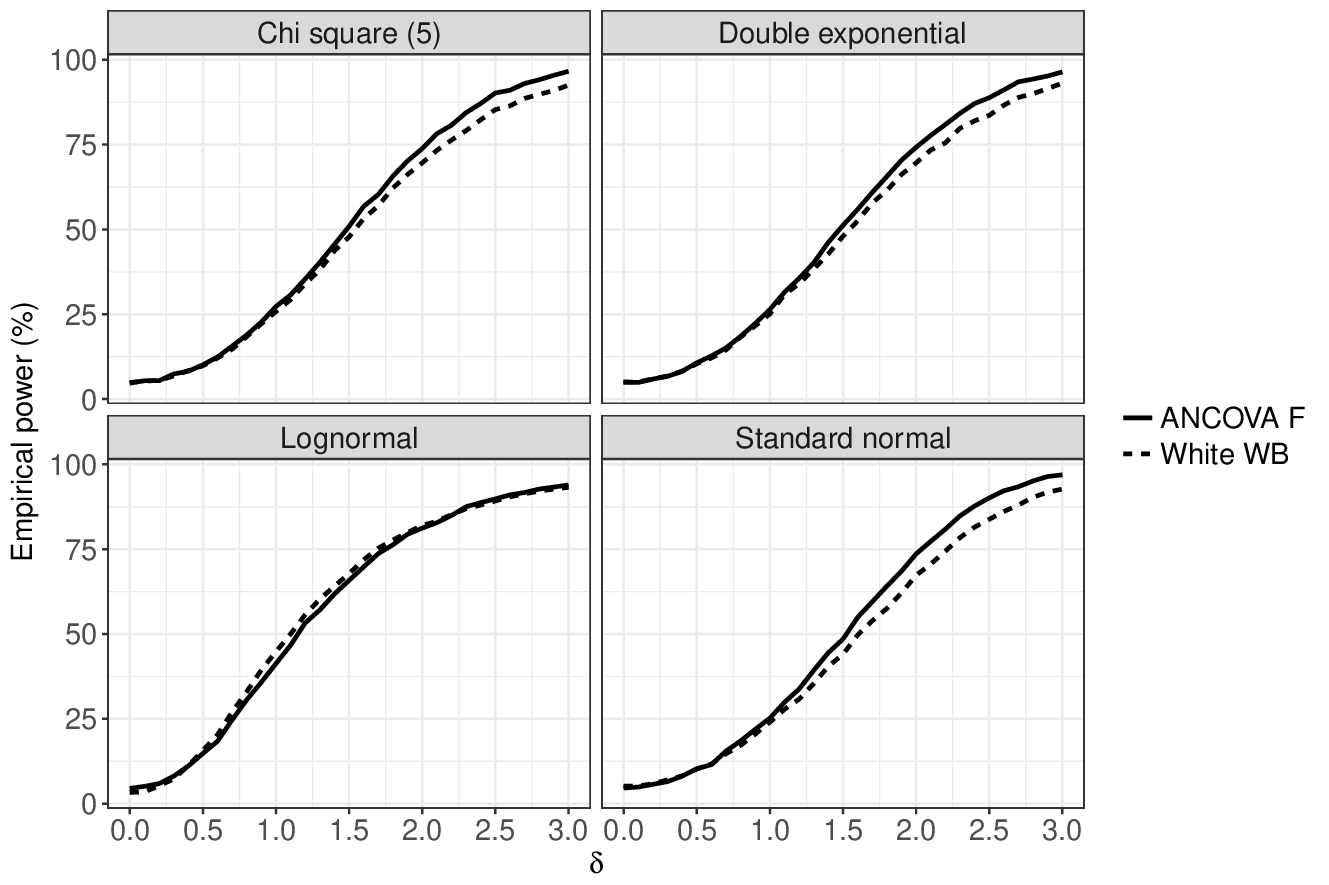}
\caption{Empirical power for the ANCOVA F test (solid) and the wild bootstrap version of the White-ANCOVA test (HC4 version; dashed). Data were generated for two groups with $\mu_1=0$, $\mu_2 = \delta$, $\sigma_1^2 = \sigma_2^2 = 1$, $n_1 = n_2 = 15$. The values of the first covariate were equally spaced between $-10$ and $10$, whereas the values of the second covariate were generated from a standard lognormal distribution.\label{FigurePowerRandLognormal}}
\end{figure}

\begin{figure}[t]

\centering
\includegraphics[width = 16cm, scale = 1]{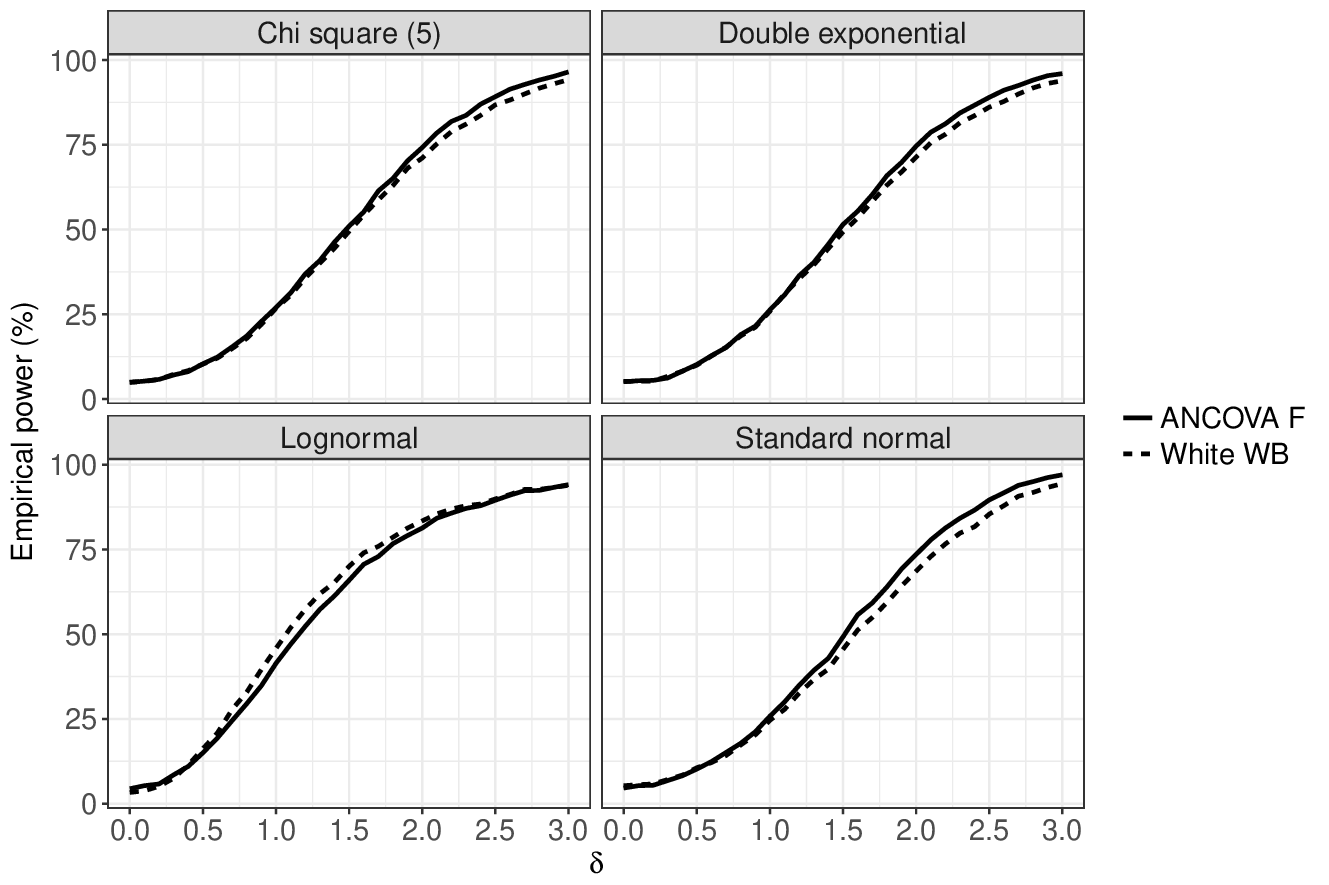}
\caption{Empirical power for the ANCOVA F test (solid) and the wild bootstrap version of the White-ANCOVA test (HC4 version; dashed). Data were generated for two groups with $\mu_1=0$, $\mu_2 = \delta$, $\sigma_1^2 = \sigma_2^2 = 1$, $n_1 = n_2 = 15$. The values of the first covariate were equally spaced between $-10$ and $10$, whereas the values of the second covariate were generated from a Poisson distribution with parameter $\lambda = 5$. \label{FigurePowerRandPoi5}}
\end{figure}

In order to assess the finite-sample performance of our proposed method in case of random covariates, we also considered the following scenarios: We left all simulation settings unchanged (for a detailed description, see Section \ref{Simulations}), but replaced the values of the covariate $\VecFormat{z}_2$ by random samples from 
\begin{itemize}
\item the uniform distribution on $[0,10]$, 
\item the standard normal distribution,
\item the standard lognormal distribution, 
\item and the $Poisson(\lambda = 5)$ distribution,
\end{itemize}
respectively. For ease of illustration, we only considered the variance scenarios I:  $\sigma_1^2 = \ldots = \sigma_4^2 = 1$ and II: $\sigma_i^2 = i$, $i \in \{1,2,3,4\}$. The results are displayed in Tables \ref{TableTypeIrandunif}-\ref{TableTypeIrandpoisson} as well as in Figures \ref{FigurePowerRandUnif}-\ref{FigurePowerRandPoi5}, where for all settings, the data generating process was repeated $10\,000$ times, and for each simulation run, $5\,000$ wild bootstrap samples were generated.

\begin{table}[h!b]
\centering

\caption{Empirical type I error rates (in $\%$) for the ANCOVA F test (F), the White-ANCOVA test (W), and
its wild bootstrap version (WB), where the latter two tests are based on the HC4 covariance matrix estimator. ${\bf n_1} = (40,40,40,40)$, ${\bf n_2} = (15,15,15,15)$, ${\bf n_3} = (5,5,5,5)$, ${\bf n_4}= (5,10,20,25)$, ${\bf n_5} = (25,20,10,5)$. The values of the first covariate were equally spaced between $-10$ and $10$, whereas the values of the second covariate were generated from a uniform distribution on $[0,10]$. } 
\vspace{5mm}
\begin{tabular}{*{14}{r}}
\hline
\hspace{0.2cm} & \hspace{0.2cm} & \multicolumn{3}{c}{\textbf{Normal}} & \multicolumn{3}{c}{\textbf{Lognormal}} & \multicolumn{3}{c}{\textbf{Double exp.}} & \multicolumn{3}{c}{{\bf Chi square(5)}}\\
\cline{3-5}\cline{6-8}\cline{9-11}\cline{12-14}
\textbf{Var}& \textbf{N}& \textbf{F}& \textbf{W}& \textbf{WB}& \textbf{F}& \textbf{W}& \textbf{WB}& \textbf{F}& \textbf{W}& \textbf{WB}& \textbf{F}& \textbf{W}& \textbf{WB}\\
\hline
$I$ & ${\bf n}_1$ & $5.0$ & $6.1$ & $5.1$ & $4.4$ & $5.0$ & $5.3$ & $4.6$ & $5.7$ & $4.9$ & $5.1$ & $6.2$ & $5.3$ \\
&${\bf n}_2$ & $5.0$ & $8.2$ & $5.2$ & $4.0$ & $5.1$ & $4.1$ & $4.9$ & $7.5$ & $4.8$ & $5.0$ & $7.4$ & $4.8$ \\
&${\bf n}_3$ & $5.0$ & $15.2$ & $5.6$ & $3.8$ & $8.3$ & $2.6$ & $4.9$ & $14.0$ & $5.0$ & $4.5$ & $13.9$ & $4.9$ \\
&${\bf n}_4$ & $5.2$ & $8.3$ & $5.2$ & $4.9$ & $5.8$ & $4.4$ & $4.8$ & $7.7$ & $5.0$ & $4.9$ & $7.3$ & $4.7$ \\
&${\bf n}_5$ & $4.9$ & $7.6$ & $4.7$ & $4.6$ & $5.1$ & $3.9$ & $5.1$ & $7.7$ & $5.0$ & $4.8$ & $7.5$ & $4.8$ \\
$II$&${\bf n}_1$ & $5.4$ & $5.9$ & $4.9$ & $4.6$ & $5.5$ & $6.0$ & $4.9$ & $5.8$ & $5.0$ & $5.3$ & $6.5$ & $5.7$ \\
&${\bf n}_2$ & $5.3$ & $8.3$ & $5.4$ & $4.3$ & $5.6$ & $4.8$ & $5.1$ & $7.5$ & $5.0$ & $5.1$ & $7.3$ & $4.8$ \\
&${\bf n}_3$ & $5.6$ & $14.8$ & $5.1$ & $4.0$ & $8.8$ & $2.7$ & $4.9$ & $13.9$ & $5.0$ & $4.9$ & $13.9$ & $5.1$ \\
&${\bf n}_4$ & $2.9$ & $7.6$ & $5.3$ & $2.7$ & $5.2$ & $4.3$ & $3.1$ & $7.3$ & $5.2$ & $2.6$ & $6.7$ & $4.5$ \\
&${\bf n}_5$ & $10.9$ & $8.0$ & $4.8$ & $8.3$ & $7.3$ & $5.7$ & $10.9$ & $8.5$ & $5.4$ & $10.6$ & $8.6$ & $5.3$ \\
\hline
\label{TableTypeIrandunif}
\end{tabular}  

\end{table}

\begin{table}[h!b]
\centering
\caption{Empirical type I error rates (in $\%$) for the ANCOVA F test (F), the White-ANCOVA test (W), and
its wild bootstrap version (WB), where the latter two tests are based on the HC4 covariance matrix estimator. ${\bf n_1} = (40,40,40,40)$, ${\bf n_2} = (15,15,15,15)$, ${\bf n_3} = (5,5,5,5)$, ${\bf n_4}= (5,10,20,25)$, ${\bf n_5} = (25,20,10,5)$. The values of the first covariate were equally spaced between $-10$ and $10$, whereas the values of the second covariate were generated from a standard normal distribution.} 
\vspace{5mm}
\begin{tabular}{*{14}{r}}
\hline
\hspace{0.2cm} & \hspace{0.2cm} & \multicolumn{3}{c}{\textbf{Normal}} & \multicolumn{3}{c}{\textbf{Lognormal}} & \multicolumn{3}{c}{\textbf{Double exp.}} & \multicolumn{3}{c}{{\bf Chi square(5)}}\\
\cline{3-5}\cline{6-8}\cline{9-11}\cline{12-14}
\textbf{Var}& \textbf{N}& \textbf{F}& \textbf{W}& \textbf{WB}& \textbf{F}& \textbf{W}& \textbf{WB}& \textbf{F}& \textbf{W}& \textbf{WB}& \textbf{F}& \textbf{W}& \textbf{WB}\\
\hline
$I$ & ${\bf n}_1$ & $5.2$ & $6.1$ & $5.0$ & $4.6$ & $5.2$ & $5.6$ & $4.8$ & $6.4$ & $5.4$ & $5.1$ & $5.7$ & $4.9$\\
& ${\bf n}_2$ & $4.9$ & $7.7$ & $5.0$ & $3.8$ & $5.0$ & $4.0$ & $5.1$ & $8.0$ & $5.3$ & $4.7$ & $7.5$ & $5.0$\\
& ${\bf n}_3$ &$4.8$ & $15.3$ & $5.6$ & $3.5$ & $8.1$ & $2.4$ & $4.9$ & $13.9$ & $4.8$ & $4.8$ & $13.7$ & $4.7$\\
& ${\bf n}_4$ &$4.8$ & $8.0$ & $4.9$ & $4.5$ & $5.2$ & $4.1$ & $4.9$ & $7.3$ & $4.7$ & $5.4$ & $7.8$ & $5.1$\\
& ${\bf n}_5$ & $5.1$ & $8.2$ & $5.4$ & $4.3$ & $5.1$ & $4.1$ & $5.2$ & $7.7$ & $5.0$ & $4.9$ & $7.7$ & $5.0$\\
$II$& ${\bf n}_1$ & $5.5$ & $6.1$ & $5.1$ & $4.6$ & $5.8$ & $6.3$ & $5.2$ & $6.2$ & $5.3$ & $4.9$ & $5.8$ & $5.0$\\
	& ${\bf n}_2$ & $5.2$ & $7.9$ & $5.2$ & $4.1$ & $5.5$ & $4.5$ & $5.3$ & $8.0$ & $5.3$ & $5.0$ & $7.6$ & $5.1$\\
	& ${\bf n}_3$ & $5.2$ & $15.0$ & $5.5$ & $3.8$ & $8.6$ & $2.6$ & $5.0$ & $13.9$ & $4.8$ & $5.0$ & $13.9$ & $4.8$\\
	& ${\bf n}_4$ & $2.8$ & $7.5$ & $5.0$ & $2.9$ & $4.8$ & $4.2$ & $2.9$ & $6.6$ & $4.6$ & $3.1$ & $7.3$ & $5.0$\\
	& ${\bf n}_5$ & $11.0$ & $8.5$ & $5.4$ & $7.8$ & $7.4$ & $5.8$ & $10.6$ & $8.4$ & $5.5$ & $11.0$ & $8.9$ & $5.8$\\
\hline
\end{tabular}  
\label{TableTypeIrandnormal}
\end{table}

\begin{table}[h!b]
\centering
\caption{Empirical type I error rates (in $\%$) for the ANCOVA F test (F), the White-ANCOVA test (W), and
its wild bootstrap version (WB), where the latter two tests are based on the HC4 covariance matrix estimator. ${\bf n_1} = (40,40,40,40)$, ${\bf n_2} = (15,15,15,15)$, ${\bf n_3} = (5,5,5,5)$, ${\bf n_4}= (5,10,20,25)$, ${\bf n_5} = (25,20,10,5)$. The values of the first covariate were equally spaced between $-10$ and $10$, whereas the values of the second covariate were generated from a standard lognormal distribution.} 
\vspace{5mm}
\begin{tabular}{*{14}{r}}
\hline
\hspace{0.1cm} & \hspace{0.1cm} & \multicolumn{3}{c}{\textbf{Normal}} & \multicolumn{3}{c}{\textbf{Lognormal}} & \multicolumn{3}{c}{\textbf{Double exp.}} & \multicolumn{3}{c}{{\bf Chi square(5)}}\\
\cline{3-5}\cline{6-8}\cline{9-11}\cline{12-14}
\textbf{Var}& \textbf{N}& \textbf{F}& \textbf{W}& \textbf{WB}& \textbf{F}& \textbf{W}& \textbf{WB}& \textbf{F}& \textbf{W}& \textbf{WB}& \textbf{F}& \textbf{W}& \textbf{WB}\\
\hline
$I$ & ${\bf n}_1$ & $5.1$ & $5.8$ & $5.0$ & $4.7$ & $5.0$ & $5.6$ & $4.8$ & $6.2$ & $5.5$ & $4.9$ & $5.5$ & $4.9$\\
& ${\bf n}_2$ & $5.0$ & $7.2$ & $5.0$ & $3.8$ & $4.6$ & $4.0$ & $5.3$ & $7.2$ & $5.3$ & $4.8$ & $6.9$ & $4.8$\\
& ${\bf n}_3$ & $4.7$ & $14.1$ & $5.6$ & $3.6$ & $7.6$ & $2.5$ & $5.0$ & $13.2$ & $5.0$ & $4.7$ & $12.7$ & $4.9$\\
& ${\bf n}_4$ & $4.7$ & $7.3$ & $4.9$ & $4.5$ & $4.9$ & $4.0$ & $4.8$ & $6.8$ & $4.7$ & $5.3$ & $7.1$ & $4.6$\\
& ${\bf n}_5$ & $5.1$ & $7.5$ & $5.5$ & $4.3$ & $4.8$ & $4.1$ & $5.1$ & $7.2$ & $5.1$ & $4.9$ & $7.2$ & $5.1$\\
$II$& ${\bf n}_1$ & $5.5$ & $5.7$ & $5.0$ & $4.6$ & $5.4$ & $6.3$ & $5.2$ & $6.1$ & $5.5$ & $4.9$ & $5.5$ & $4.9$\\
	& ${\bf n}_2$ & $5.2$ & $7.1$ & $4.8$ & $4.1$ & $4.9$ & $4.6$ & $5.4$ & $7.2$ & $5.3$ & $5.0$ & $6.9$ & $4.9$\\
	& ${\bf n}_3$ & $5.0$ & $13.9$ & $5.6$ & $3.6$ & $8.0$ & $2.5$ & $4.9$ & $13.6$ & $5.0$ & $5.0$ & $13.1$ & $5.0$\\
	& ${\bf n}_4$ & $2.9$ & $6.8$ & $5.1$ & $3.0$ & $4.8$ & $4.4$ & $2.9$ & $6.1$ & $4.6$ & $3.0$ & $6.5$ & $4.8$\\
	& ${\bf n}_5$ & $11.0$ & $7.9$ & $5.5$ & $7.8$ & $6.8$ & $5.9$ & $10.8$ & $8.0$ & $5.5$ & $11.3$ & $8.5$ & $5.8$\\
\hline
\end{tabular}  
\label{TableTypeIrandlognormal}
\end{table}

\begin{table}[h!t]
\centering

\caption{Empirical type I error rates (in $\%$) for the ANCOVA F test (F), the White-ANCOVA test (W), and
its wild bootstrap version (WB), where the latter two tests are based on the HC4 covariance matrix estimator. ${\bf n_1} = (40,40,40,40)$, ${\bf n_2} = (15,15,15,15)$, ${\bf n_3} = (5,5,5,5)$, ${\bf n_4}= (5,10,20,25)$, ${\bf n_5} = (25,20,10,5)$. The values of the first covariate were equally spaced between $-10$ and $10$, whereas the values of the second covariate were generated from a Poisson distribution with parameter $\lambda = 5$.} 
\vspace{5mm}
\begin{tabular}{*{14}{r}}
\hline
\hspace{0.1cm} & \hspace{0.1cm} & \multicolumn{3}{c}{\textbf{Normal}} & \multicolumn{3}{c}{\textbf{Lognormal}} & \multicolumn{3}{c}{\textbf{Double exp.}} & \multicolumn{3}{c}{{\bf Chi square(5)}}\\
\cline{3-5}\cline{6-8}\cline{9-11}\cline{12-14}
\textbf{Var}& \textbf{N}& \textbf{F}& \textbf{W}& \textbf{WB}& \textbf{F}& \textbf{W}& \textbf{WB}& \textbf{F}& \textbf{W}& \textbf{WB}& \textbf{F}& \textbf{W}& \textbf{WB}\\
\hline
$I$ & ${\bf n}_1$ & $5.0$ & $5.9$ & $5.0$ & $4.4$ & $4.9$ & $5.4$ & $4.7$ & $5.7$ & $5.0$ & $5.1$ & $6.2$ & $5.4$ \\
&${\bf n}_2$ & $4.9$ & $8.1$ & $5.1$ & $3.9$ & $5.0$ & $4.1$ & $4.8$ & $7.4$ & $4.9$ & $4.9$ & $7.3$ & $4.7$ \\
&${\bf n}_3$ & $4.9$ & $15.0$ & $5.5$ & $3.7$ & $8.6$ & $2.8$ & $4.8$ & $14.0$ & $5.0$ & $4.5$ & $13.9$ & $4.8$ \\
&${\bf n}_4$ & $5.2$ & $8.3$ & $5.3$ & $4.9$ & $5.8$ & $4.4$ & $4.9$ & $7.7$ & $5.0$ & $4.9$ & $7.2$ & $4.8$ \\
&${\bf n}_5$ & $4.9$ & $7.5$ & $4.8$ & $4.5$ & $5.1$ & $4.0$ & $5.1$ & $7.6$ & $4.9$ & $4.8$ & $7.4$ & $4.9$ \\
$II$&${\bf n}_1$ & $5.4$ & $6.0$ & $4.8$ & $4.6$ & $5.4$ & $6.0$ & $4.9$ & $5.7$ & $5.0$ & $5.3$ & $6.5$ & $5.8$ \\
&${\bf n}_2$ & $5.3$ & $8.1$ & $5.4$ & $4.3$ & $5.6$ & $4.8$ & $5.1$ & $7.4$ & $5.1$ & $5.0$ & $7.2$ & $4.7$ \\
&${\bf n}_3$ & $5.5$ & $14.9$ & $5.1$ & $4.0$ & $8.8$ & $3.0$ & $4.8$ & $13.9$ & $5.1$ & $4.8$ & $13.9$ & $5.0$ \\
&${\bf n}_4$ & $2.9$ & $7.7$ & $5.2$ & $2.7$ & $5.2$ & $4.3$ & $3.0$ & $7.3$ & $5.2$ & $2.5$ & $6.6$ & $4.6$ \\
&${\bf n}_5$ & $10.8$ & $8.0$ & $4.7$ & $8.3$ & $7.2$ & $5.7$ & $10.8$ & $8.4$ & $5.4$ & $10.7$ & $8.8$ & $5.4$ \\
\hline
\label{TableTypeIrandpoisson}
\end{tabular}  

\end{table}

\pagebreak
\newpage

\clearpage
\pagebreak

\section{Further simulation results: More severe heteroskedasticity}

\vspace{1cm}

All simulation parameters were the same as described in Section \ref{Simulations}, but the variances of groups 1 to 4 were set to $\sigma_1^2 = 1$, $\sigma_2^2 = 5$, $\sigma_3^2 = 10$ and $\sigma_4^2 = 20$, respectively. The data generating process was repeated $10\,000$ times for all settings. For each simulation run, $5\,000$ wild bootstrap samples were generated.

\begin{table}[h!b]
\centering
\caption{Empirical type I error rates (in $\%$) for the ANCOVA F test (F), the White-ANCOVA test (W), and
its wild bootstrap version (WB), where the latter two tests are based on the HC4 covariance matrix estimator. ${\bf n_1} = (40,40,40,40)$, ${\bf n_2} = (15,15,15,15)$, ${\bf n_3} = (5,5,5,5)$, ${\bf n_4}= (5,10,20,25)$, ${\bf n_5} = (25,20,10,5)$. The values of the first covariate were equally spaced between $-10$ and $10$. For the second vector of covariate values, the first and the second half of the components were equally spaced in $[0,5]$ and $[-2,-1]$, respectively, sorted in descending order. The group variances were set to $\sigma_1^2 = 1, \sigma_2^2 = 5, \sigma_3^3 = 10, \sigma_4^2 = 20$. } 
\vspace{5mm}
\begin{tabular}{*{13}{r}}
\hline
 \hspace{0.2cm} & \multicolumn{3}{c}{\textbf{Normal}} & \multicolumn{3}{c}{\textbf{Lognormal}} & \multicolumn{3}{c}{\textbf{Double exp.}} & \multicolumn{3}{c}{{\bf Chi square(5)}}\\
\cline{2-4}\cline{5-7}\cline{8-10}\cline{11-13}
\textbf{N}& \textbf{F}& \textbf{W}& \textbf{WB}& \textbf{F}& \textbf{W}& \textbf{WB}& \textbf{F}& \textbf{W}& \textbf{WB}& \textbf{F}& \textbf{W}& \textbf{WB}\\
\hline
${\bf n}_1$ & $4.5$ & $6.4$ & $4.9$ & $5.0$ & $6.6$ & $7.0$ & $4.9$ & $6.6$ & $5.6$ & $4.7$ & $6.4$ & $5.3$\\
${\bf n}_2$ & $4.8$ & $8.9$ & $5.3$ & $4.7$ & $8.0$ & $6.0$ & $5.1$ & $8.7$ & $5.4$ & $4.9$ & $8.5$ & $5.2$\\
${\bf n}_3$ &$5.2$ & $15.6$ & $5.7$ & $4.4$ & $11.9$ & $3.9$ & $5.4$ & $15.8$ & $5.9$ & $5.2$ & $15.5$ & $5.8$\\
${\bf n}_4$ &$2.6$ & $8.2$ & $5.0$ & $2.7$  & $9.0$ & $7.1$ & $2.3$ & $8.2$ & $5.3$ & $2.6$ & $8.7$ & $5.8$\\
${\bf n}_5$ & $16.4$ & $8.2$ & $5.0$ & $13.4$& $7.1$ & $4.9$ & $16.6$ & $8.6$ & $5.3$ & $16.3$ & $8.7$ & $5.2$\\
\hline
\label{TableTypeIseverehet}
\end{tabular}  

\end{table}

\clearpage
\pagebreak

\section{Further simulation results: Highly unbalanced group / very small total sample sizes}

\vspace{1cm}

We considered a setting with $a=2$ groups of sizes ${\bf n_1} = (5,5)$, ${\bf n_2} = (5,10)$, ${\bf n_3} = (5,20)$, ${\bf n_4}= (5,40)$, ${\bf n_5} = (5,80)$, respectively. One homoskedastic (Var I: $\sigma_1^2 = \sigma_2^2 = 1$) and two heteroskedastic (Var II: $\sigma_1^2 = 1, \sigma_2^2 = 4$, Var III: $\sigma_1^2 = 4$, $\sigma_2^2 = 1$) settings were examined. All other specifications were the same as described in Section \ref{Simulations}. The data generating process was repeated $10\,000$ times. For each simulation run, $5\,000$ wild bootstrap samples were generated.

\begin{table}[h!b]
\centering
\caption{Empirical type I error rates (in $\%$) for the ANCOVA F test (F), the White-ANCOVA test (W), and
its wild bootstrap version (WB), where the latter two tests are based on the HC4 covariance matrix estimator. ${\bf n_1} = (5,5)$, ${\bf n_2} = (5,10)$, ${\bf n_3} = (5,20)$, ${\bf n_4}= (5,40)$, ${\bf n_5} = (5,80)$. Var I: $\sigma_1^2 = \sigma_2^2 = 1$, Var II: $\sigma_1^2 = 1, \sigma_2^2 = 4$, Var III: $\sigma_1^2 = 4$, $\sigma_2^2 = 1$. All further specifications were the same as described in Section \ref{Simulations}.} 
\vspace{5mm}
\begin{tabular}{*{14}{r}}
\hline
\hspace{0.2cm} & \hspace{0.2cm} & \multicolumn{3}{c}{\textbf{Normal}} & \multicolumn{3}{c}{\textbf{Lognormal}} & \multicolumn{3}{c}{\textbf{Double exp.}} & \multicolumn{3}{c}{{\bf Chi square(5)}}\\
\cline{3-5}\cline{6-8}\cline{9-11}\cline{12-14}
\textbf{Var}& \textbf{N}& \textbf{F}& \textbf{W}& \textbf{WB}& \textbf{F}& \textbf{W}& \textbf{WB}& \textbf{F}& \textbf{W}& \textbf{WB}& \textbf{F}& \textbf{W}& \textbf{WB}\\
\hline
$I$ & ${\bf n}_1$ & $4.9$ & $9.2$ & $8.4$ & $4.7$ & $6.5$ & $6.2$ & $5.2$ & $8.8$ & $8.0$ & $5.3$ & $8.7$ & $8.4$ \\
&${\bf n}_2$ & $4.9$ & $9.6$ & $9.1$ & $5.9$ & $6.2$ & $6.4$ & $4.9$ & $8.7$ & $8.3$ & $5.2$ & $8.5$ & $8.5$ \\
&${\bf n}_3$ & $5.0$ & $7.8$ & $6.4$ & $5.0$ & $5.2$ & $4.4$ & $4.7$ & $6.7$ & $5.4$ & $5.1$ & $7.3$ & $6.0$ \\
&${\bf n}_4$ & $5.2$ & $6.1$ & $5.3$ & $5.3$ & $3.9$ & $3.2$ & $5.1$ & $5.4$ & $4.8$ & $4.8$ & $5.5$ & $4.6$ \\
&${\bf n}_5$ & $5.2$ & $4.2$ & $5.2$ & $5.4$ & $5.5$ & $6.7$ & $5.1$ & $4.5$ & $5.8$ & $4.8$ & $4.5$ & $5.9$ \\
$II$&${\bf n}_1$ & $5.2$ & $10.5$ & $9.3$ & $5.9$ & $8.5$ & $7.5$ & $5.3$ & $9.9$ & $8.6$ & $5.4$ & $9.7$ & $9.3$ \\
&${\bf n}_2$ & $1.7$ & $8.9$ & $8.9$ & $3.5$ & $7.2$ & $7.1$ & $1.8$ & $8.2$ & $8.1$ & $2.2$ & $7.9$ & $8.1$ \\
&${\bf n}_3$ & $4.1$ & $8.6$ & $7.6$ & $4.4$ & $9.4$ & $8.9$ & $4.0$ & $7.9$ & $7.0$ & $4.1$ & $8.7$ & $7.7$ \\
&${\bf n}_4$ & $1.6$ & $5.9$ & $5.2$ & $2.6$ & $5.6$ & $5.5$ & $2.1$ & $5.9$ & $5.3$ & $2.0$ & $5.5$ & $4.9$ \\
&${\bf n}_5$ & $0.6$ & $3.9$ & $4.8$ & $1.6$ & $2.7$ & $3.6$ & $0.6$ & $3.7$ & $4.7$ & $0.6$ & $3.7$ & $4.4$ \\
$III$&${\bf n}_1$ & $6.5$ & $11.0$ & $10.2$ & $6.7$ & $9.3$ & $8.3$ & $7.1$ & $10.7$ & $9.4$ & $7.3$ & $10.9$ & $10.4$ \\
&${\bf n}_2$ & $10.9$ & $9.6$ & $9.2$ & $10.4$ & $8.4$ & $8.2$ & $10.8$ & $9.0$ & $8.6$ & $10.5$ & $9.0$ & $9.1$ \\
&${\bf n}_3$ & $7.6$ & $7.7$ & $6.2$ & $8.0$ & $6.4$ & $5.1$ & $7.3$ & $7.1$ & $5.6$ & $8.2$ & $8.0$ & $6.2$ \\
&${\bf n}_4$ & $16.3$ & $7.1$ & $5.6$ & $14.5$ & $10.5$ & $10.3$ & $15.8$ & $7.2$ & $6.3$ & $15.2$ & $8.2$ & $7.2$ \\
&${\bf n}_5$ & $23.5$ & $5.1$ & $5.3$ & $19.1$ & $12.5$ & $14.6$ & $23.0$ & $6.0$ & $6.8$ & $22.3$ & $6.7$ & $7.5$ \\
\hline
\label{TableTypeIsamplesizes}
\end{tabular}  

\end{table}

\clearpage
\pagebreak




\section{Further simulation results: Bias and variance}

\vspace{1cm}

Since in the linear model under consideration, the vector of parameter estimators and its wild bootstrap counterpart are  (unconditionally) unbiased, there would have been little use in conducting simulations in this case. Nevertheless, the variances of those parameter estimators as well as the bias and the variance of the HC4 version of White's covariance matrix estimator and its wild bootstrap analogon might depend on the sample sizes as well as other specifications. Therefore, we conducted simulations for the very same settings as the ones that are described in Section \ref{Simulations}, yet focusing on the aforementioned properties of the estimators. For ease of presentation, we only considered variance scenarios I and II. At first, we simulated the (unconditional) empirical bias and variance of each subject-specific variance estimator and averaged over all subjects, then. Likewise, we examined the average empirical variance of the estimated adjusted means (the regression part of the parameter vector was of no interest in the present manuscript). More formally, the average empirical variance of the estimated adjusted means $V_{AM}$ and the average empirical bias $B_W$ and variance $V_W$ of the subject-specific White covariance matrix estimator were calculated according to  
\begin{align*}
V_{AM} &= \frac{1}{a}\sum_{i=1}^{a}\frac{1}{n_{sim}-1}\sum_{k=1}^{n_{sim}}(\hat{\mu}_{i,k} - \bar{\hat{\mu}}_{i.})^2, \\
B_{W} &= \frac{1}{N}\sum_{i=1}^{a}\sum_{j=1}^{n_i}\frac{1}{n_{sim}}\sum_{k=1}^{n_{sim}}(\hat{\sigma}_{ij,k}^2 - \sigma_{i}^2),\\
V_{W} &= \frac{1}{N}\sum_{i=1}^{a}\sum_{j=1}^{n_i}\frac{1}{n_{sim}-1}\sum_{k=1}^{n_{sim}}(\hat{\sigma}_{ij,k}^2 - \bar{\hat{\sigma}}_{ij.}^2)^2,
\end{align*}
where $n_{sim}$ denotes the number of simulation runs, bars/dots indicate averaging over all simulation runs, and $N = n_1 + \ldots + n_a$. In order to calculate the respective unconditional empirical variances and bias  for the bootstrap-based estimators, we used the empirical counterparts of the law of iterated expectations (\textit{i.e.}, $E[X] = E[E[X|Y]]$) and the equation $Var[X] = E[Var[X|Y]] + Var[E[X|Y]]$ for random variables $X,Y$. Doing so, we obtained
\begin{align*}
V_{AM}^{\ast} &= \frac{1}{a}\sum_{i=1}^{a}\left(\frac{1}{n_{sim}}\sum_{k=1}^{n_{sim}}\frac{1}{n_{boot}-1}\sum_{l=1}^{n_{boot}}(\hat{\mu}_{i,kl}^{\ast} - \bar{\hat{\mu}}_{i,k.}^{\ast})^2 + \frac{1}{n_{sim}-1}\sum_{k=1}^{n_{sim}}( \bar{\hat{\mu}}_{i,k.}^{\ast} - \bar{\hat{\mu}}_{i..}^{\ast})^2\right),\\
B_{W}^{\ast} &= \frac{1}{N}\sum_{i=1}^{a}\sum_{j=1}^{n_i}\left(\frac{1}{n_{sim}}\sum_{k=1}^{n_{sim}}\frac{1}{n_{boot}}\sum_{l=1}^{n_{boot}}(\hat{\sigma}_{ij,kl}^{\ast 2} - \sigma_{i}^2)\right),\\
V_{W}^{\ast} &= \frac{1}{N}\sum_{i=1}^{a}\sum_{j=1}^{n_i}\left(\frac{1}{n_{sim}}\sum_{k=1}^{n_{sim}}\frac{1}{n_{boot}-1}\sum_{l=1}^{n_{boot}}(\hat{\sigma}_{ij,kl}^{\ast 2} - \bar{\hat{\sigma}}_{ij,k.}^{\ast 2})^2 + \frac{1}{n_{sim}-1}\sum_{k=1}^{n_{sim}}( \bar{\hat{\sigma}}_{ij,k.}^{\ast 2} - \bar{\hat{\sigma}}_{ij..}^{\ast 2})^2\right),
\end{align*}   
where the dot/bar notation indicates averaging over the simulation or bootstrap runs, depending on the context, and $n_{boot}$ denotes the number of bootstrap iterations per simulation run.  
 The results are reported in Table \ref{TableBiasVarOriginal} and Table \ref{TableBiasVarBootstrap}.  \\
Summing up, the variance of the estimated adjusted means seemed to depend neither on the error distribution nor on the estimation method (i.e., White HC4 vs. White HC4 wild bootstrap). Needless to say that the variance increased with decreasing sample sizes. Moreover, the variances seemed to be larger in the heteroskedastic setting, which might well be due to the fact that the original group variances were larger than in the homoskedastic setting. The bias of the White HC4 estimators was again very similar for both methods, indicating that the variances were overestimated on average. In line with the theoretical considerations from the proofs (see Section 1 of this Online Appendix), the bias decreased with increasing sample sizes. Moreover, the average bias appeared to be smaller for positive pairing ($n_4$ in variance setting II) compared to negative pairing ($n_5$ in variance setting II). Finally, some differences between the White estimator and its bootstrap counterpart were seen in the variances of the respective estimators, with the latter being superior to the former in most scenarios. Overall, the variances were larger in heteroskedastic settings. Apparently, both estimators showed  remarkably increased variances for lognormal errors compared to the other distributions under consideration, which may be explained by the large skewness. So, all in all, the performance of the estimators might be somewhat suboptimal in this case, which corresponds to the slightly conservative behavior that has been found in the simulation results reported in Section \ref{Simulations}.

\begin{table}[h!b]
\centering

\caption{Average empirical variance of the estimated adjusted means ($V_{AM}$) as well as average empirical bias ($B_{W}$) and average empirical variance ($V_{W}$) of the White HC4 covariance matrix estimator. All specifications were the same as the ones used for the type I error rate simulations described in Section \ref{Simulations}. The data generating process was repeated $10\,000$ times.} 
\vspace{5mm}
\begin{tabular}{*{14}{r}}
\hline
\hspace{0.2cm} & \hspace{0.2cm} &\multicolumn{3}{c}{\textbf{Normal}} & \multicolumn{3}{c}{\textbf{Lognormal}} & \multicolumn{3}{c}{\textbf{Double exp.}} & \multicolumn{3}{c}{{\bf Chi square(5)}}\\
\cline{3-5}\cline{6-8}\cline{9-11}\cline{12-14}
\textbf{Var}& \textbf{N}& \textbf{$V_{AM}$}& \textbf{$B_{W}$}& \textbf{$V_{W}$}&  \textbf{$V_{AM}$}& \textbf{$B_{W}$}& \textbf{$V_{W}$}& \textbf{$V_{AM}$}& \textbf{$B_{W}$}& \textbf{$V_{W}$}&  \textbf{$V_{AM}$}& \textbf{$B_{W}$}& \textbf{$V_{W}$}\\
\hline
$I$ & ${\bf n}_1$ & $0.17$ & $0.00$ & $2.02$ & $0.17$ & $0.01$ & $87.08$ & $0.17$ & $0.00$ & $4.25$ & $0.17$ & $0.00$ & $4.28$ \\
&${\bf n}_2$ & $0.46$ & $0.01$ & $2.06$ & $0.46$ & $0.01$ & $72.75$ & $0.46$ & $0.01$ & $4.01$ & $0.46$ & $0.02$ & $4.11$ \\
&${\bf n}_3$ & $1.31$ & $0.04$ & $2.20$ & $1.33$ & $0.06$ & $64.69$ & $1.30$ & $0.03$ & $3.40$ & $1.31$ & $0.04$ & $3.53$ \\
&${\bf n}_4$ & $0.40$ & $0.03$ & $2.12$ & $0.39$ & $0.02$ & $84.19$ & $0.40$ & $0.03$ & $4.11$ & $0.39$ & $0.02$ & $4.13$ \\
&${\bf n}_5$ & $0.55$ & $0.03$ & $2.14$ & $0.55$ & $0.05$ & $110.46$ & $0.56$ & $0.03$ & $4.20$ & $0.55$ & $0.03$ & $4.27$ \\
$II$ &${\bf n}_1$ & $0.35$ & $0.01$ & $15.11$ & $0.36$ & $0.03$ & $747.88$ & $0.36$ & $0.01$ & $31.74$ & $0.36$ & $0.01$ & $31.77$ \\
&${\bf n}_2$ & $0.97$ & $0.03$ & $15.39$ & $0.96$ & $0.02$ & $572.67$ & $0.97$ & $0.03$ & $29.82$ & $0.97$ & $0.04$ & $31.27$ \\
&${\bf n}_3$ & $2.92$ & $0.12$ & $16.49$ & $2.96$ & $0.16$ & $320.74$ & $2.88$ & $0.07$ & $25.23$ & $2.91$ & $0.10$ & $25.70$ \\
&${\bf n}_4$ & $1.08$ & $0.03$ & $21.06$ & $1.06$ & $0.00$ & $810.51$ & $1.08$ & $0.04$ & $41.50$ & $1.07$ & $0.03$ & $42.59$ \\
&${\bf n}_5$ & $1.26$ & $0.12$ & $11.19$ & $1.27$ & $0.17$ & $476.82$ & $1.29$ & $0.12$ & $20.88$ & $1.28$ & $0.12$ & $21.08$ \\
\hline
\label{TableBiasVarOriginal}
\end{tabular}  

\end{table}

\begin{table}[h!b]
\centering

\caption{Average unconditional empirical variance of the bootstrap versions of the estimated adjusted means ($V_{AM}^{\ast}$) as well as average unconditional empirical bias ($B_{W}^{\ast}$) and average unconditional empirical variance ($V_{W}^{\ast}$) of the bootstrap version of the White HC4 covariance matrix estimator. All specifications were the same as the ones used for the type I error rate simulations described in Section \ref{Simulations}. The data generating process was repeated $10\,000$ times. Within each simulation run, $5\,000$ bootstrap iterations were performed.} 
\vspace{5mm}
\begin{tabular}{*{14}{r}}
\hline
\hspace{0.2cm} & \hspace{0.2cm} &\multicolumn{3}{c}{\textbf{Normal}} & \multicolumn{3}{c}{\textbf{Lognormal}} & \multicolumn{3}{c}{\textbf{Double exp.}} & \multicolumn{3}{c}{{\bf Chi square(5)}}\\
\cline{3-5}\cline{6-8}\cline{9-11}\cline{12-14}
\textbf{Var}& \textbf{N}& \textbf{$V_{AM}^{\ast}$}& \textbf{$B_{W}^{\ast}$}& \textbf{$V_{W}^{\ast}$}&  \textbf{$V_{AM}^{\ast}$}& \textbf{$B_{W}^{\ast}$}& \textbf{$V_{W}^{\ast}$}& \textbf{$V_{AM}^{\ast}$}& \textbf{$B_{W}^{\ast}$}& \textbf{$V_{W}^{\ast}$}&  \textbf{$V_{AM}^{\ast}$}& \textbf{$B_{W}^{\ast}$}& \textbf{$V_{W}^{\ast}$}\\
\hline
$I$ & "${\bf n}_1$ & $0.17$ & $0.00$ & $2.02$ & $0.18$ & $0.01$ & $80.63$ & $0.17$ & $0.00$ & $4.08$ & $0.17$ & $0.00$ & $4.11$ \\
&${\bf n}_2$ & $0.45$ & $0.01$ & $2.07$ & $0.45$ & $0.01$ & $60.31$ & $0.45$ & $0.01$ & $3.68$ & $0.46$ & $0.02$ & $3.76$ \\
&${\bf n}_3$ & $1.32$ & $0.04$ & $2.55$ & $1.33$ & $0.06$ & $57.23$ & $1.32$ & $0.03$ & $3.54$ & $1.32$ & $0.04$ & $3.65$ \\
&${\bf n}_4$ & $0.40$ & $0.03$ & $2.15$ & $0.39$ & $0.02$ & $71.85$ & $0.40$ & $0.03$ & $3.81$ & $0.39$ & $0.02$ & $3.84$ \\
&${\bf n}_5$ & $0.54$ & $0.03$ & $2.17$ & $0.55$ & $0.05$ & $93.06$ & $0.54$ & $0.03$ & $3.88$ & $0.54$ & $0.03$ & $3.95$ \\
$II$&${\bf n}_1$ & $0.36$ & $0.01$ & $15.10$ & $0.36$ & $0.03$ & $691.89$ & $0.36$ & $0.01$ & $30.52$ & $0.36$ & $0.01$ & $30.54$ \\
&${\bf n}_2$ & $0.96$ & $0.03$ & $15.45$ & $0.95$ & $0.02$ & $470.67$ & $0.96$ & $0.03$ & $27.36$ & $0.96$ & $0.04$ & $28.52$ \\
&${\bf n}_3$ & $2.94$ & $0.12$ & $19.02$ & $2.98$ & $0.16$ & $282.75$ & $2.91$ & $0.07$ & $26.26$ & $2.92$ & $0.10$ & $26.65$ \\
&${\bf n}_4$ & $1.08$ & $0.03$ & $21.18$ & $1.05$ & $0.00$ & $696.79$ & $1.08$ & $0.04$ & $38.51$ & $1.06$ & $0.03$ & $39.55$ \\
&${\bf n}_5$ & $1.26$ & $0.12$ & $11.45$ & $1.29$ & $0.17$ & $386.71$ & $1.26$ & $0.12$ & $19.24$ & $1.26$ & $0.12$ & $19.44$ \\
\hline
\label{TableBiasVarBootstrap}
\end{tabular}  

\end{table}

\clearpage

\bibliography{IntroBibl}

\begin{thebibliography}{}

\bibitem[\protect\astroncite{Adams et~al.}{1985}]{Ada}
Adams, K., Brown, G., and Grant, I. (1985).
\newblock Analysis of covariance as a remedy for demographic mismatch of
  research subject groups: Some sobering simulations.
\newblock {\em Journal of Clinical and Experimental Neuropsychology},
  7(4):445--462.

\bibitem[\protect\astroncite{Ashford and Brown}{1969}]{Ash}
Ashford, J. and Brown, S. (1969).
\newblock Generalised covariance analysis with unequal error variances.
\newblock {\em Biometrics}, 25(4):715--724.

\bibitem[\protect\astroncite{Barton et~al.}{2013}]{Bar}
Barton, S., Crozier, S., Lillycrop, K., Godfrey, K., and Inskip, H. (2013).
\newblock Correction of unexpected distributions of p values from analysis of
  whole genome arrays by rectifying violation of statistical assumptions.
\newblock {\em BMC Genomics}, 14:161.

\bibitem[\protect\astroncite{Bathke and Brunner}{2003}]{Bat}
Bathke, A. and Brunner, E. (2003).
\newblock A nonparametric alternative to analysis of covariance.
\newblock In Akritas, M. and Politis, D., editors, {\em Recent Advantages and
  Trends in Nonparametric Statistics}, pages 109--120. Elsevier, Amsterdam.

\bibitem[\protect\astroncite{Berman and Greenhouse}{1992}]{Ber}
Berman, N. and Greenhouse, S. (1992).
\newblock Adjusting for demographic covariates by the analysis of covariance.
\newblock {\em Journal of Clinical and Experimental Neuropsychology},
  14(6):981--982.

\bibitem[\protect\astroncite{Beyersmann et~al.}{2013}]{Bey}
Beyersmann, J., Di~Termini, S., and Pauly, M. (2013).
\newblock Weak convergence of the wild bootstrap for the {A}alen-{J}ohansen
  estimator of the cumulative incidence function of a competing risk.
\newblock {\em Scandinavian Journal of Statistics}, 78:387--402.

\bibitem[\protect\astroncite{Bossa et~al.}{2011}]{Bos}
Bossa, M., Zacur, E., Olmos, S., and {Alzheimer's Disease Neuroimaging
  Initiative} (2011).
\newblock Statistical analysis of relative pose information of subcortical
  nuclei: Application on {ADNI} data.
\newblock {\em Neuroimage}, 55(3):999--1008.

\bibitem[\protect\astroncite{Cameron et~al.}{2008}]{Cam08}
Cameron, A., Gelbach, J., and Miller, D. (2008).
\newblock Bootstrap-based improvements for inference with clustered errors.
\newblock {\em The Review of Economics and Statistics}, 90(3):414--427.

\bibitem[\protect\astroncite{Chausse et~al.}{2016}]{Cha}
Chausse, P., Liu, J., and Luta, G. (2016).
\newblock A simulation-based comparison of covariate adjustment methods for the
  analysis of randomized controlled trials.
\newblock {\em International Journal of Environmental Research and Public
  Health}, 13:414.

\bibitem[\protect\astroncite{Craggs et~al.}{2006}]{Cra06}
Craggs, M., Balasubramaniam, A., Chung, E., and Emmanuel, A. (2006).
\newblock Aberrant reflexes and function of the pelvic organs following spinal
  cord injury in man.
\newblock {\em Autonomic Neuroscience: Basic \& Clinical}, 126--127:355--70.

\bibitem[\protect\astroncite{Cribari-Neto}{2004}]{Cri}
Cribari-Neto, F. (2004).
\newblock Asymptotic inference under heteroskedasticity of unknown form.
\newblock {\em Computational Statistics and Data Analysis}, 45:215--233.

\bibitem[\protect\astroncite{Cruz and Cruz}{2011}]{Cru11}
Cruz, C. and Cruz, F. (2011).
\newblock Spinal cord injury and bladder dysfunction: new ideas about an old
  problem.
\newblock {\em The Scientific World Journal}, 11:214--34.

\bibitem[\protect\astroncite{Davidson and Flachaire}{2008}]{Dav}
Davidson, R. and Flachaire, E. (2008).
\newblock The wild bootstrap, tamed at last.
\newblock {\em Journal of Econometrics}, 146(1):162--169.

\bibitem[\protect\astroncite{{European Medicines Agency}}{2015}]{Ema15}
{European Medicines Agency} (2015).
\newblock Guideline on adjustment for baseline covariates in clinical trials
  {EMA/CHMP/295050/2013}.

\bibitem[\protect\astroncite{Fan and Zhang}{2017}]{Fan17}
Fan, C. and Zhang, D. (2017).
\newblock Rank repeated measures analysis of covariance.
\newblock {\em Communications in Statistics - Theory and Methods},
  46(3):1158--1183.

\bibitem[\protect\astroncite{Glass et~al.}{1972}]{glass1972consequences}
Glass, G., Peckham, P., and Sanders, J. (1972).
\newblock Consequences of failure to meet assumptions underlying the fixed
  effects analyses of variance and covariance.
\newblock {\em Review of educational research}, 42(3):237--288.

\bibitem[\protect\astroncite{Hayes and Cai}{2007}]{Hay}
Hayes, A. and Cai, L. (2007).
\newblock Using heteroskedasticity-consistent standard error estimators in
  {OLS} regression: An introduction and software implementation.
\newblock {\em Behavior Research Methods}, 39(4):709--722.

\bibitem[\protect\astroncite{Huitema}{2011}]{Hui}
Huitema, B. (2011).
\newblock {\em The Analysis of Covariance and Alternatives: Statistical Methods
  for Experiments, Quasi-Experiments, and Single-Case Studies}.
\newblock Wiley, New York.

\bibitem[\protect\astroncite{Judkins and Porter}{2016}]{Jud}
Judkins, D. and Porter, K. (2016).
\newblock Robustness of ordinary least squares in randomized clinical trials.
\newblock {\em Statistics in Medicine}, 35:1763--1773.

\bibitem[\protect\astroncite{Keselman et~al.}{1998}]{keselman1998statistical}
Keselman, H., Huberty, C., Lix, L., Olejnik, S., Cribbie, R., Donahue, B.,
  Kowalchuk, R., Lowman, L., Petoskey, M., Keselman, J., and Levin, J. (1998).
\newblock Statistical practices of educational researchers: An analysis of
  their {ANOVA}, {MANOVA}, and {ANCOVA} analyses.
\newblock {\em Review of Educational Research}, 68(3):350--386.

\bibitem[\protect\astroncite{Kimura}{1990}]{Kim}
Kimura, D. (1990).
\newblock Testing nonlinear regression parameters under heteroscedastic,
  normally distributed errors.
\newblock {\em Biometrics}, 46(3):697--708.

\bibitem[\protect\astroncite{Koch et~al.}{1998}]{Koc}
Koch, G., Tangen, C., Jung, J., and Amara, I. (1998).
\newblock Issues for covariance analysis of dichotomous and ordered categorical
  data from randomized clinical trials and non-parametric strategies for
  addressing them.
\newblock {\em Statistics in Medicine}, 17:1863--1892.

\bibitem[\protect\astroncite{Lesaffre and Senn}{2003}]{lesaffre2003note}
Lesaffre, E. and Senn, S. (2003).
\newblock A note on non-parametric {ANCOVA} for covariate adjustment in
  randomized clinical trials.
\newblock {\em Statistics in Medicine}, 22(23):3583--3596.

\bibitem[\protect\astroncite{Liu}{1988}]{Liu}
Liu, R. (1988).
\newblock Bootstrap procedures under some non-i.i.d. models.
\newblock {\em The Annals Of Statistics}, 16(4):1696--1708.

\bibitem[\protect\astroncite{Lu}{2014}]{Lu14}
Lu, K. (2014).
\newblock An efficient analysis of covariance model for crossover thorough {QT}
  studies with period-specific pre-dose baselines.
\newblock {\em Pharmaceutical Statistics}, 13(6):388--396.

\bibitem[\protect\astroncite{MacKinnon}{2012}]{Mac10}
MacKinnon, J. (2012).
\newblock Thirty years of heteroskedasticity-robust inference.
\newblock {\em Queen's Economic Department Working Paper 1268}.

\bibitem[\protect\astroncite{MacKinnon and White}{1985}]{Mac}
MacKinnon, J. and White, H. (1985).
\newblock Some heteroskedasticity-consistent covariance matrix estimators with
  improved finite sample properties.
\newblock {\em Journal of Econometrics}, 29:305--325.

\bibitem[\protect\astroncite{Mammen}{1993}]{Mam}
Mammen, E. (1993).
\newblock Bootstrap and wild bootstrap for high dimensional linear models.
\newblock {\em The Annals of Statistics}, 21(1):255--285.

\bibitem[\protect\astroncite{Misra}{1996}]{Mis}
Misra, R. (1996).
\newblock A multivariate procedure for comparing mean vectors for populations
  with unequal regression coefficient and residual covariance matrices.
\newblock {\em Biometrical Journal}, 38:415--424.

\bibitem[\protect\astroncite{Mitsui et~al.}{2014}]{Mit14}
Mitsui, T., Murray, M., and Nonomura, K. (2014).
\newblock Lower urinary tract function in spinal cord-injured rats: midthoracic
  contusion versus transsection.
\newblock {\em Spinal Cord}, 52(9):658--61.

\bibitem[\protect\astroncite{Owen and Froman}{1998}]{Owe}
Owen, S. and Froman, R. (1998).
\newblock Uses and abuses of the analysis of covariance.
\newblock {\em Research in Nursing \& Health}, 21:557--562.

\bibitem[\protect\astroncite{Pocock et~al.}{2002}]{Poc}
Pocock, S., Assmann, S., Enos, L., and Kasten, L. (2002).
\newblock Subgroup analysis, covariate adjustment and baseline comparisons in
  clinical trial reporting: Current practice and problems.
\newblock {\em Statistics in Medicine}, 21:2917--2930.

\bibitem[\protect\astroncite{{R Development Core Team}}{2017}]{Rco17}
{R Development Core Team} (2017).
\newblock R: A language and environment for statistical computing.
\newblock R Foundation for Statistical Computing, Vienna, Austria.
  \url{http://www.R-project.org}.

\bibitem[\protect\astroncite{Ravishanker and Dey}{2002}]{Rav}
Ravishanker, N. and Dey, D. (2002).
\newblock {\em A First Course In Linear Model Theory}.
\newblock Chapman \& Hall/CRC, New York.

\bibitem[\protect\astroncite{Sadooghi-Alvandi and Jafari}{2013}]{Sad}
Sadooghi-Alvandi, S. and Jafari, A. (2013).
\newblock A parametric bootstrap approach for one-way {ANCOVA} with unequal
  variances.
\newblock {\em Communications in Statistics: Theory and Methods},
  42(14):2473--2498.

\bibitem[\protect\astroncite{Senn}{2006}]{Sen}
Senn, S. (2006).
\newblock Change from baseline and analysis of covariance revisited.
\newblock {\em Statistics in Medicine}, 25:4334--4344.

\bibitem[\protect\astroncite{Tangen and Koch}{2001}]{Tan}
Tangen, C. and Koch, G. (2001).
\newblock Non-parametric analysis of covariance for confirmatory randomized
  clinical trials to evaluate dose-response relationships.
\newblock {\em Statistics in Medicine}, 20:2585--2607.

\bibitem[\protect\astroncite{Thas et~al.}{2012}]{Tha}
Thas, O., {De Neve}, J., Clement, L., and Ottoy, J. (2012).
\newblock Probabilistic index models.
\newblock {\em Journal of the Royal Statistical Society: Series B (Statistical
  Methodology)}, 74(4):623--671.

\bibitem[\protect\astroncite{Tsiatis et~al.}{2008}]{Tsi}
Tsiatis, A., Davidian, M., Zhang, M., and Lu, X. (2008).
\newblock Covariate adjustment for two-sample treatment comparisons in
  randomized clinical trials: A principled yet flexible approach.
\newblock {\em Statistics in Medicine}, 27:4658--4677.

\bibitem[\protect\astroncite{Van~der Vaart}{2007}]{Van}
Van~der Vaart, A. (2007).
\newblock {\em Asymptotic Statistics}.
\newblock Cambridge University Press, New York, 8th edition.

\bibitem[\protect\astroncite{White}{1980}]{Whi}
White, H. (1980).
\newblock A heteroskedasticity-consistent covariance matrix estimator and a
  direct test for heteroskedasticity.
\newblock {\em Econometrica}, 48:817--838.

\bibitem[\protect\astroncite{Wu}{1986}]{Wu}
Wu, C. (1986).
\newblock Jackknife, bootstrap and other resampling methods in regression
  analysis.
\newblock {\em The Annals of Statistics}, 14(4):1261--1295.

\bibitem[\protect\astroncite{Zhu et~al.}{2008}]{Zhu}
Zhu, T., Liu, X., Connelly, P., and Zhong, J. (2008).
\newblock An optimized wild bootstrap method for evaluation of measurement
  uncertainties of {DTI}-derived parameters in human brain.
\newblock {\em Neuroimage}, 40:1144--1156.

\end{thebibliography}

\end{document}